\documentclass[preprint,12pt,authoryear]{elsarticle}

\usepackage{amssymb}
\usepackage{amsmath}

\usepackage{bm}
\usepackage[unicode=true,
    psdextra,
    colorlinks=true,
    pdfstartview=FitV,
    linkcolor=blue,
    citecolor=blue,
    urlcolor=blue,
    anchorcolor = blue]{hyperref}
\usepackage{physics}
\usepackage{mlmodern}
\usepackage{booktabs}
\usepackage{multirow}

\usepackage[textwidth=6in, top=1in, bottom=1.2in]{geometry}

\journal{Journal of Non-Newtonian Fluid Mechanics}

\begin{document}

\begin{frontmatter}



\title{On the deformation of a shear thinning viscoelastic drop in a steady electric field} 


\author[a]{Sarika Shivaji Bangar} 
\author[a]{Gaurav Tomar}

\affiliation{organization={Department of Mechanical Engineering, Indian Institute of Science},
            addressline={}, 
            city={Bengaluru},
            postcode={560012}, 
            state={Karnataka},
            country={India}}

\begin{abstract}
The deformation of viscoelastic drops under electric fields plays a crucial role in applications such as microfluidics, inkjet printing, and electrohydrodynamic manipulation of complex fluids. This study examines the deformation and breakup dynamics of a linear Phan-Thien–Tanner (LPTT) drop subjected to a uniform electric field using numerical simulations performed with the open-source solver Basilisk. Representative combinations of conductivity ratio ($\sigma_r$) and permittivity ratio ($\epsilon_r$) are chosen from six characteristic regions of the ($\sigma_r$, $\epsilon_r$) phase space, $PR_A^+$, $PR_B^+$, $PR_A^-$, $PR_B^-$, $OB^+$, and $OB^-$.
In regions where the first- and second-order deformation coefficients have the same sign ($PR_A^-$, $PR_B^-$, $OB^+$), the LPTT drops exhibit deformation dynamics that negligibley deviate from the Newtonian behavior. In the $PR_A^+$ region, drops deform into prolate spheroidal shapes below a critical electric capillary number and transition to stable multi-lobed shapes or breakup beyond this threshold. Increasing elasticity of drop opposes the deformation, thereby reducing deformation and increasing critical $Ca_E$ with the Deborah number ($De$).
In the $PR_B^+$ region, drops form prolate shapes below critical $Ca_E$ and develop conical ends above it. The steady-state deformation exhibits a non-monotonic dependence on $De$, increasing at low $De$ and decreasing at higher values. A similar non-monotonic variation is also observed in critical $Ca_E$.
In the $OB^-$ region, LPTT drops attain oblate shapes below critical $Ca_E$ and undergo breakup beyond it. The deformation magnitude shows a non-monotonic variation with $De$, increasing initially and decreasing at higher elasticity.
Overall, these findings highlight the complex interplay between viscoelasticity and electric stresses in determining drop deformation, stability and morphology, offering insights for controlled manipulation of viscoelastic drops in electrohydrodynamic systems.
\end{abstract}




\end{frontmatter}


\section{Introduction}
Understanding the deformation dynamics of droplets in electric fields is fundamental to a wide range of applications, including inkjet printing (\cite{basaran2013nonstandard}, \cite{lau2017ink}), oil de-emulsification (\cite{eow2002electrostatic}, \cite{alvarado2010enhanced}, \cite{zhang2011application}), electrospraying and atomization (\cite{kelly1984electrostatic} ,\cite{law2018electrostatic}), electrostatic coating (\cite{hines1966electrostatic}), electric propulsion (\cite{moreau2015electrohydrodynamic}, \cite{huh2019numerical}), and droplet manipulation in microfluidics (\cite{laser2004review}, \cite{stone2004engineering}, \cite{phan2025demand}). These processes also shed light on natural phenomena such as rain electrification and droplet breakup in thunderstorms (\cite{simpson1909electricity}, \cite{wilson1921iii}, \cite{blanchard1963electrification}).

A dielectric drop subjected to an electric field typically elongates into a prolate spheroid at low field strengths and undergoes breakup at high fields (\cite{o1953distortion}, \cite{taylor1964disintegration}). \cite{allan1962particle} attributed this deformation to the balance between interfacial electric stresses and surface tension, while \cite{o1953distortion} arrived at similar conclusions using an energy-based approach. To explain the experimentally observed oblate shapes, \cite{o1957electric} identified fluid conductivity as a key parameter governing prolate and oblate forms. Building on these insights, \cite{taylor1966studies} developed the classical leaky-dielectric model (LDM), demonstrating that tangential electric stresses induce internal circulation and dictate the deformation mode based on conductivity and permittivity ratios. This framework was later refined by \cite{melcher1969electrohydrodynamics}, extended to alternating fields by \cite{torza1971electrohydrodynamic}, and further corrected for higher-order effects by \cite{ajayi1978note}. However, experimental observations by \cite{torza1971electrohydrodynamic} revealed deformations larger than theoretical predictions, and higher-order corrections failed to reconcile this difference. \cite{saville1997electrohydrodynamics} has provided a comprehensive review of the leaky dielectric model.

The LDM and its extensions have since been used to analyze diverse electrohydrodynamic (EHD) phenomena. Surfactants were shown to amplify deformation, particularly for prolate shapes (\cite{ha1995effects}), while the extended LDM (ELDM) (\cite{bentenitis2005droplet}) captured hysteresis and continuous deformation at high fields. \cite{zabarankin2013liquid} modeled moderate deformations using a spheroidal approximation, while \cite{nganguia2013equilibrium} extended this to surfactant-laden drops. \cite{deshmukh2013deformation} further showed that in quadrupole fields, deformation is dominated by the fourth-order Legendre mode with complex internal circulation.
Drop electrorotation, first examined by \cite{he2013electrorotation} and later generalized by \cite{yariv2016electrohydrodynamic}, was linked to surface charge convection and Quincke-type instabilities. The Taylor–Melcher framework has since been applied to a wide range of symmetry-breaking instabilities, including Quincke rotation, equatorial streaming and pattern formation \cite{vlahovska2016electrohydrodynamic}, \cite{vlahovska2019electrohydrodynamics}. The charge convection is shown to strongly affect the breakup dynamics (\cite{sengupta2017role}).

\cite{sozou1972electrohydrodynamics} first analyzed transient drop deformation by incorporating fluid inertia into the governing equations. Using a quasi-steady approximation, \cite{moriya1986deformation} studied time-dependent deformation in weak fields. \cite{esmaeeli2011transient} derived a closed-form solution quantifying the roles of normal and tangential electric stresses. \cite{zhang2013transient} extended the Taylor–Melcher model to include finite charge relaxation time for spheroidal drops. \cite{lanauze2013influence} further accounted for inertia and showed that deformation overshoot depends on the Ohnesorge and Saville numbers. \cite{lanauze2015nonlinear} examined nonlinear effects with surface charge convection, revealing transient prolate-to-oblate transitions. \cite{esmaeeli2020transient} performed 3D simulations showing that drop dynamics range from monotonic to oscillatory depending on viscous–inertial timescales.

Experimental efforts have provided further insights into EHD deformation and breakup. \cite{nishiwaki1988deformation} experimentally related drop retardation time to fluid properties, finding good agreement with theory for viscous drops but lower deformation for water. \cite{vizika1992electrohydrodynamic} observed improved, though not perfect, consistency with leaky-dielectric predictions under steady and oscillatory fields. \cite{karp2024electrohydrodynamic} used PIV and shadowgraphy to show close agreement with the model for small deformations. \cite{brosseau2017streaming} reported equatorial streaming in low-viscosity drops under strong DC fields, driven by electric shear stresses that concentrate flow and trigger fluid ejection at the equator.

Numerical studies have significantly advanced the understanding of Newtonian EHD behavior. \cite{miksis1981shape} used boundary integrals to show that dielectric drops form conical ends above a critical permittivity ratio, while \cite{sherwood1988breakup} extended this to leaky-dielectric drops, predicting breakup modes such as tip streaming and blob formation. His results linked high permittivity ratios to pointed-end drops and high conductivities to blob division, though the analysis was limited to prolate deformations. \cite{feng1996computational} solved the full EHD problem via finite element method, identifying a critical electric field for instability and showing strong dependence on conductivity and viscosity ratios; \cite{feng1999electrohydrodynamic} later found charge convection reduces oblate but enhances prolate deformation. \cite{lac2007axisymmetric} analyzed stability across wide property ratios, including oblate modes. \cite{dubash2007behaviour} captured steady and breakup regimes for inviscid conducting drops, while \cite{karyappa2014breakup} classified viscosity-dependent breakup identifying three axisymmetric prior to breakup (ASPB) modes (lobes, pointed, non-pointed ends) that transform into non-axisymmetric shape at breakup (NSAB) modes (charged disintegration, regular jet, open jet). 
More recent studies by \cite{wagoner2021electrohydrodynamics} and \cite{wang2024lattice}, linked equatorial streaming and fingering instabilities to coupled surface charge convection and diffusion under strong DC fields.

Several computational techniques have been developed for simulating multiphase EHD flows. \cite{zhang20052d} introduced a 2D multicomponent LBM, while \cite{tomar2007two} used a coupled level-set/VOF method and proposed weighted harmonic mean interpolation for property smoothing. \cite{hua2008numerical} combined front-tracking and VOF methods for various electric models, and \cite{paknemat2012numerical} used a level-set method with a finite difference based ghost fluid approach to capture tip-streaming and breakup. \cite{nganguia2015immersed} employed the Immersed Interface Method, and \cite{das2017electrohydrodynamics} developed a 3D boundary element method including surface charge convection and Quinke rotation. Later, \cite{liu2019phase} and \cite{wang2024lattice} advanced LBM-based multiphase solvers coupling Allen–Cahn, Poisson, and Navier–Stokes equations for charge and field evolution.

Many natural and industrial fluids, such as polymer solutions and biological fluids, are viscoelastic, exhibiting coupled viscous and elastic responses that lead to normal stress differences, stress relaxation, and strain hardening. These effects alter interfacial dynamics, delaying breakup and modifying the critical capillary number.
\citet{ramaswamy1999deformation}, \citet{hooper2001transient}, and \citet{aggarwal2007deformation} showed that viscoelasticity generally reduces drop deformation, enhances shape recovery, and increases breakup resistance in extensional and shear flows. Despite the prevalence of viscoelastic drops in numerous natural and industrial systems, literature on their electrohydrodynamics is limited. 
\citet{ha1999deformation} investigated the effect of viscoelasticity on drop deformation and stability in an electric field using a second-order fluid model. Building on this, \citet{ha2000deformation} analyzed Newtonian and non-Newtonian conducting drops, showing that both shear-independent and shear-rate-dependent viscoelastic fluids exhibit reduced deformation and improved stability. They further found that elasticity in the surrounding phase stabilizes drops at low viscosity ratios but can induce instability at higher ratios. Later, \citet{lima2014numerical} numerically studied a Giesekus drop and showed that increasing polymer relaxation time suppresses deformation through elastic resistance while also lowering effective viscosity via shear thinning. More recently, \citet{zhao2025electrohydrodynamic} examined viscoelastic drops in combined electric and shear flows, finding that elasticity diminishes deformation and rotation, though its impact remains secondary to electrohydrodynamic stresses. Recently, \citet{das2026effect} developed an analytical approximation for the deformation of an Oldroyd-B drop in an electric field in the small Deborah number regime and conducted numerical simulations to study the drop dynamics at higher Deborah numbers. In a recent study, \citet{bangar2026large} investigated the large deformation of an Oldroyd-B drop subjected to an electric field. Although the Oldroyd-B model captures the elastic nature of viscoelastic fluids, it does not account for shear-thinning behavior and permits infinite extension of polymer molecules, which is physically unrealistic.

Motivated by these limitations, the present study investigates the deformation of a drop described by the linear Phan-Thien–Tanner (LPTT) model subjected to a steady uniform electric field. The LPTT model incorporates finite polymer extensibility and shear-thinning behavior, providing a more realistic representation of viscoelastic fluid behavior.
Pairs of conductivity ($\sigma_r$) and permittivity ($\epsilon_r$) ratios are selected based on the deformation–circulation phase map (\cite{lac2007axisymmetric}, \cite{bangar2026large}) obtained from small-deformation analysis (\cite{das2026effect}). The governing equations and boundary conditions are described in \S\ref{Sec_Formualtion}, followed by the numerical methodology and grid independence study in \S\ref{Sec_Numerics}. The results are discussed in \S\ref{Sec_results}, and the key conclusions of the study are summarized in \S\ref{Sec_Conclusion}.

\section{Problem Formulation}\label{Sec_Formualtion}
\begin{figure}[ht!]
	\centering\includegraphics{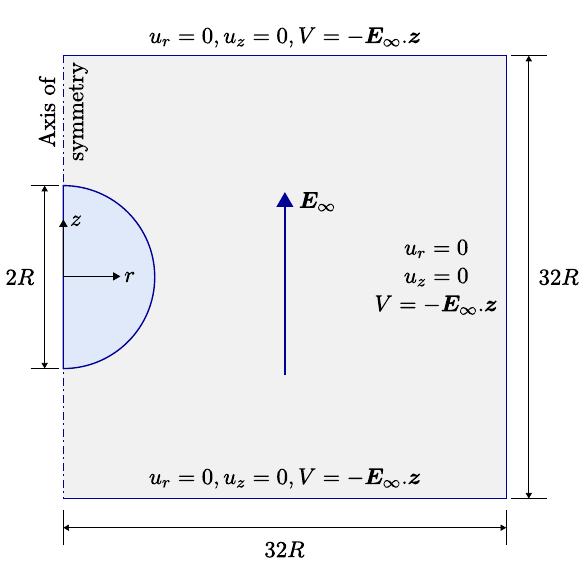}
	\caption{Schematic of the problem. LPTT drop of radius $R$ is subjected to an external electric field, $\bm{E}_\infty$, aligned along the axis of symmetry. The domain size is set to $32R \times 32R$ to minimize the boundary effects.}
	\label{Fig_schematic}
\end{figure}
A viscoelastic drop of radius $R$, suspended in a Newtonian fluid, is subjected to a uniform, steady electric field $\bm{E}_{\infty}$. Subscripts $i$ and $e$ denote the drop and ambient phases, respectively. Each phase is characterized by its density $\rho$, viscosity $\mu$, permittivity $\epsilon$, and conductivity $\sigma$. The ambient fluid is Newtonian, while the drop is modeled using the linear Phan--Thien--Tanner (PTT) constitutive equation with relaxation time $\lambda_i$, solvent and polymeric viscosities $\mu_{i_s}$ and $\mu_{i_p}$, and extensibility parameter $\varepsilon$. Axisymmetric simulations are performed using the open-source solver \href{https://www.basilisk.fr/}{Basilisk}  (\cite{popinet2015quadtree}), and the system schematic is shown in \autoref{Fig_schematic}.

\subsection{Governing Equations}
The incompressible fluid flow is governed by the mass and momentum conservation equations:
\begin{equation}
    \bm{\nabla}\cdot\bm{u} = 0, \quad \rho \frac{D \bm{u}}{Dt} = -\bm{\nabla} p + \bm{\nabla} \cdot \bm{\tau} + \bm{F},
\end{equation}
where $\bm{u}$ is the velocity, $\rho$ the density, $p$ the pressure, $\bm{\tau}$ the stress tensor, and $\bm{F}$ the body force. The stress tensor is related to the velocity gradient through a constitutive relation. In the presence of an electric field $\bm{E}$, the body force arises from the Maxwell stress tensor $\bm{\tau}_E$ as
\begin{equation}
    \bm{F} = \bm{\nabla} \cdot \bm{\tau}_E, \quad \bm{\tau}_E = \epsilon \left(\bm{E}\bm{E} - \frac{1}{2} (\bm{E}\cdot\bm{E}) \bm{I} \right),
\end{equation}
where $\epsilon$ is the permittivity.  

The viscoelastic drop is modeled using the linear Phan–Thien–Tanner (LPTT) model, where the total stress is
\begin{equation}
    \bm{\tau} = \bm{\tau}_s + \bm{\tau}_p,
\end{equation}
with the solvent stress $\bm{\tau}_s = 2 \mu_s \bm{D}$, where $\bm{D} = \frac{1}{2}(\bm{\nabla u} + (\bm{\nabla u})^T)$, and the polymeric stress tensor evolves according to
\begin{equation}
    \bm{\tau}_p \left(1 + \frac{\varepsilon \lambda}{\mu_p} \mathrm{tr}(\bm{\tau}_p)\right) + \lambda \stackrel{\nabla}{\bm{\tau}}_p = \mu_p (\bm{\nabla u} + (\bm{\nabla u})^T),
\end{equation}
where $\lambda$ is the relaxation time, $\mu_p = \mu - \mu_s$ the polymeric viscosity, $\varepsilon$ the extensibility parameter, and $\stackrel{\nabla}{\bm{\tau}}_p$ denotes the upper-convected derivative.  

Magnetic effects are neglected as the characteristic magnetic time $t_M = \mu_M \sigma L^2$ is much smaller than the electric relaxation time $t_E = \epsilon / \sigma$. The electric field is governed by Gauss's and Faraday's laws:
\begin{equation}
    \bm{\nabla} \cdot (\epsilon \bm{E}) = q, \quad \bm{\nabla} \times \bm{E} = 0, 
\end{equation}
where, $q$ denotes the volumetric charge density. Electric field can be represented in terms of a gradient of electric potential ($\bm{E} = -\bm{\nabla} V$), 
leading to the following equation for the electric potential,
\begin{equation}
    \bm{\nabla} \cdot (\epsilon \bm{\nabla} V) = -q.
\end{equation}
The bulk free charge conservation is expressed as
\begin{equation}
    \frac{\partial q}{\partial t} + \bm{\nabla} \cdot \bm{J} = 0, \quad \bm{J} = \sigma \bm{E} + q \bm{u},
\end{equation}
where the first term in $\bm{J}$ represents Ohmic conduction and the second term represents charge convection. Solving these equations numerically provides the electric field distribution, which is used to compute $\bm{\tau}_E$ and the resulting electric body force.

\subsection{Boundary Conditions}
A schematic of the simulation domain is shown in \autoref{Fig_schematic}, with the origin of the $(r,z)$ coordinate system at the drop center. The domain extends from $-L/2$ to $L/2$ in $z$ and from $0$ to $L$ in $r$. Far from the drop, the velocity vanishes:  
\begin{subequations}
\begin{align}
    u_r(r,-L/2) &= u_z(r,-L/2) = 0 \\
    u_r(r,L/2) &= u_z(r,L/2) = 0, \\
    u_r(L,z) &= u_z(L,z) = 0.
\end{align}
\end{subequations}
The imposed uniform electric field $\bm{E}_\infty$ is represented by the potential $V$:
\begin{equation}
    \bm{E}_\infty = -\bm{\nabla} V, \quad V = -E_\infty z.
\end{equation}
Accordingly, the electric potential satisfies
\begin{subequations}
\begin{align}
    V(r,-L/2) = V(r,L/2) = V(L,z) = -E_\infty z.
\end{align}
\end{subequations}

All polymeric stress components are assigned Neumann boundary conditions far from the drop, which introduces only local errors (\citet{alves2021numerical}). Symmetry is imposed at the axis $r=0$:
\begin{subequations}
\begin{align}
    &u_r(0,z) = 0, \quad \pdv{u_z}{r}\Big|_{r=0} = 0 \\  &\quad \pdv{V}{r}\Big|_{r=0} = 0, \\
    &\pdv{\tau_{p_{rr}}}{r}\Big|_{r=0} = \pdv{\tau_{p_{zz}}}{r}\Big|_{r=0} = \tau_{p_{rz}}(0,z) = \pdv{\tau_{p_{\theta\theta}}}{r}\Big|_{r=0} = 0.
\end{align}
\end{subequations}

\subsection{Non-dimensional Parameters}
To simplify the analysis, we non-dimensionalize the governing equations. The drop radius $R$ is used as a characteristic length scale and the magnitude of the imposed electric field, $E_\infty$ is taken as a characteristic scale for the electric field. The velocity scale is obtained by the balance of electric and viscous stress, 
\begin{equation}
    U \sim \frac{\epsilon_e E_{\infty}^2 R}{\mu_e}.
\end{equation}
As flow is viscous dominated, stresses and pressure are non-dimensionalized by the viscous scale, $\mu_e U/R$. This viscous stress scale is equivalent to the electric stress scale $\epsilon_e E_{\infty}^2$, as the velocity scale itself is derived from their balance. Upon non-dimensionalization of the governing equations, we get the following dimensionless parameters that affect the problem:
\begin{enumerate}
    \item Permittivity ratio, $\epsilon_r = \frac{\epsilon_i}{\epsilon_e}$
    \item Conductivity ratio, $\sigma_r = \frac{\sigma_i}{\sigma_e}$
    \item Density ratio, $\rho_r = \frac{\rho_i}{\rho_e}$
    \item Viscosity ratio, $\mu_r = \frac{\mu_i}{\mu_e}$
    \item Flow Reynolds number, $Re = \frac{\rho_e U R}{\mu_e}$ 
    \item Electric capillary number, $Ca_E = \frac{\epsilon_e E_\infty^2 R}{\gamma}$, is the ratio of electric force to surface tension force with the surface tension being a resisting force and the force due to the electric field being a deforming force.
    \item Deborah number, $De = \frac{\lambda_i U}{R}$, is the ratio of the polymer relaxation time to the flow time scale.
    \item Ratio of solvent to total viscosity of the drop phase, $\beta_i = \frac{\mu_{i_s}}{\mu_i}$
    \item Extensibility parameter of LPTT fluid, $\varepsilon$
\end{enumerate}

\section{Numerical Simulations}\label{Sec_Numerics}
\subsection{Numerical Methodology}
For the axisymmetric numerical simulations, we employ the open-source solver \href{https://basilisk.fr}{Basilisk}(\cite{popinet2015quadtree}), which simulates two-phase flows using a geometric Volume of Fluid (VOF) method. The solver uses a second-order accurate time-splitting projection scheme with variables stored at cell centers. Diffusion terms are treated implicitly, while advection is discretized using the second-order Bell-Collela-Glaz (BCG) scheme. Equation for electric potential and charge conservation equation are solved by the numerical scheme given in \cite{lopez2011charge}.
The viscoelastic constitutive equation is solved using the log-conformation tensor approach as described by \cite{fattal2004constitutive}. The polymeric stress is expressed in terms of a symmetric positive definite tensor $\bm{A}$ as $\bm{\tau}_p = \frac{\mu_p}{\lambda}(\bm{A}-\bm{I})$. Introducing $\bm{\Psi} = \log \bm{A}$, its evolution equations is written as:
\begin{equation}\label{equ_Psi}
	\pdv{\bm{\Psi}}{t} + \underbrace{\bm{u}.\bm{\nabla}\bm{\Psi}}_{\text{Advection}} = \underbrace{\bm{\Omega}\bm{\Psi} - \bm{\Psi}\bm{\Omega} + 2\bm{B}}_{\text{Upper convection}} + \underbrace{\frac{1}{\lambda}\left(1 + \frac{\varepsilon \mu_p}{\lambda}(\text{Tr}(e^{\bm{\Psi}} - 3)\right)(e^{-\bm{\Psi}} - \bm{I})}_{\text{Model term}}.
\end{equation}
The evolution equation for $\bm{\Psi}$ is solved using a split scheme given by \cite{lopez2019adaptive}.
\begin{align}
	&\pdv{\bm{\Psi}}{t} = \bm{\Omega}\bm{\Psi} - \bm{\Psi}\bm{\Omega} + 2\bm{B} \label{Eqn_Psi_upperConv}\\
	&\pdv{\bm{\Psi}}{t} + \grad.(\bm{u\Psi}) = 0 \label{Eqn_Psi_Advestion}\\
	&\pdv{\bm{A}}{t} = \frac{1}{\lambda}(\bm{I} - \bm{A})\left(1 + \frac{\varepsilon\mu_p}{\lambda} \text{Tr}(\bm{A} - \bm{I})\right) \label{Eqn_Psi_Model}
\end{align}
The upper-convected derivative (\autoref{Eqn_Psi_upperConv}) is computed explicitly, advection of $\bm{\Psi}$ (\autoref{Eqn_Psi_Advestion}) is handled using the BCG scheme , and the LPTT model term (\autoref{Eqn_Psi_Model}) is integrated implicitly via the Euler method.

\begin{table}[ht!]
    \fontsize{11}{12}\selectfont
    \caption{Coordinates of the primary eddy center and minimum horizontal velocity along $x=0.5$ line and location of its occurrence for lid driven cavity flow of LPTT fluid}
    \vspace{8px}
    \centering
    \begin{tabular}{lcccccccc}
            \toprule
			LPTT           & $Re$                   & $De$                 & $\beta$              & $\varepsilon$            & $x_{eddy}$ & $y_{eddy}$ & $u_{min}$ & $y_{min}$ \\ \midrule
			\citet{dalal2016numerical} & \multirow{2}{*}{1}   & \multirow{2}{*}{1} & \multirow{2}{*}{0.3} & \multirow{2}{*}{0.25} & 0.481      & 0.789      & -0.1679   & 0.5349    \\
			Present        &                      &                    &                      &                       & 0.483      & 0.789      & -0.1676   & 0.533     \\ \midrule
			\citet{dalal2016numerical} & \multirow{2}{*}{1}   & \multirow{2}{*}{1} & \multirow{2}{*}{1/9} & \multirow{2}{*}{0.1}  & 0.453      & 0.800      & -0.1416   & 0.5116    \\
			Present        &                      &                    &                      &                       & 0.460      & 0.804      & -0.1393   & 0.514     \\ \midrule
			\citet{dalal2016numerical} & \multirow{2}{*}{100} & \multirow{2}{*}{1} & \multirow{2}{*}{1/9} & \multirow{2}{*}{0.25} & 0.778      & 0.845      & -0.1037   & 0.5387    \\
			Present        &                      &                    &                      &                       & 0.781      & 0.851      & -0.1026   & 0.541     \\ \midrule
			\citet{dalal2016numerical} & \multirow{2}{*}{100} & \multirow{2}{*}{4} & \multirow{2}{*}{1/9} & \multirow{2}{*}{0.25} & 0.783      & 0.767      & -0.0858   & 0.3363    \\
			Present        &                      &                    &                      &                       & 0.789      & 0.781      & -0.0844   & 0.338  \\ 
            \bottomrule
		\end{tabular}
    \label{Table_LPTT_lid_1}
\end{table}
\begin{table}[ht!]
    \fontsize{11}{12}\selectfont
    \caption{Minimum and maximum values of vertical velocity along the $y=0.5$ line and position of their occurrence for lid driven cavity flow of LPTT fluid}
    \vspace{8px}
    \centering
    \begin{tabular}{lcccccccc}
        \toprule
        LPTT           & $Re$                   & $De$                 & $\beta$              & $\varepsilon$            & $v_{min}$ & $x_{min}$ & $v_{max}$ & $x_{max}$ \\ \midrule
        \citet{dalal2016numerical} & \multirow{2}{*}{1}   & \multirow{2}{*}{1} & \multirow{2}{*}{0.3} & \multirow{2}{*}{0.25} & -0.1429   & 0.7950    & 0.1572    & 0.2016    \\
        Present        &                      &                    &                      &                       & -0.1426   & 0.795     & 0.1568    & 0.201     \\ \midrule
        \citet{dalal2016numerical} & \multirow{2}{*}{1}   & \multirow{2}{*}{1} & \multirow{2}{*}{1/9} & \multirow{2}{*}{0.1}  & -0.1235   & 0.7883    & 0.1484    & 0.1883    \\
        Present        &                      &                    &                      &                       & -0.1213   & 0.787     & 0.1454    & 0.189     \\ \midrule
        \citet{dalal2016numerical} & \multirow{2}{*}{100} & \multirow{2}{*}{1} & \multirow{2}{*}{1/9} & \multirow{2}{*}{0.25} & -0.1050   & 0.8559    & 0.0753    & 0.2339    \\
        Present        &                      &                    &                      &                       & -0.1021   & 0.853     & 0.0742    & 0.233     \\ \midrule
        \citet{dalal2016numerical} & \multirow{2}{*}{100} & \multirow{2}{*}{4} & \multirow{2}{*}{1/9} & \multirow{2}{*}{0.25} & -0.1301   & 0.9059    & 0.0903    & 0.2489    \\
        Present        &                      &                    &                      &                       & -0.1265   & 0.904     & 0.0887    & 0.248   \\ \bottomrule
        \end{tabular}
    \label{Table_LPTT_lid_2}
\end{table}
To validate the implementation, simulations of lid-driven cavity flow for an LPTT fluid are performed across a range of parameters. The results are compared with the numerical findings of \citet{dalal2016numerical}. \autoref{Table_LPTT_lid_1} presents a comparison of the primary eddy position and the minimum horizontal velocity along the $x=0.5$ centerline, including its location, while \autoref{Table_LPTT_lid_2} lists the minimum and maximum vertical velocities along the $y=0.5$ centerline with their respective locations. The tabulated results demonstrate good agreement with the literature, confirming the accuracy and robustness of the numerical approach.

\subsection{Grid convergence and domain independence study}
\begin{figure}[ht!]
    \centering
    \includegraphics{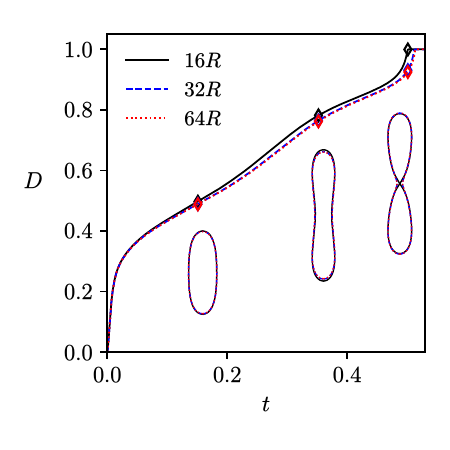}
    \caption{Deformation vs. time for various sizes of the simulation domain. Simulation parameters are $\mu_r=1$, $\rho_r=1$, $Re=1$, $\beta_i = 1/9$, $\sigma_r = 10$, $\epsilon_r=1.37$, $De=5$, $Ca_E=0.4$, $\varepsilon=0.25$. $R/\Delta x_{min}$ is taken as 256.}
    \label{Fig_domain_test}
\end{figure}
A domain independence test is conducted to determine the appropriate simulation domain size, as illustrated in \autoref{Fig_domain_test}. Numerical simulations are performed using three different domain sizes, and the evolution of the drop deformation parameter over time is examined. The figure also presents the interface shapes at three distinct time instances for each domain size. Results show that the deformation behavior for domains of size $32R$ and $64R$ closely overlap, indicating negligible difference. Consequently, a domain size of $32R$ is selected for all subsequent simulations in this study.

\begin{figure}[ht!]
	\centering\includegraphics[width=1\textwidth]{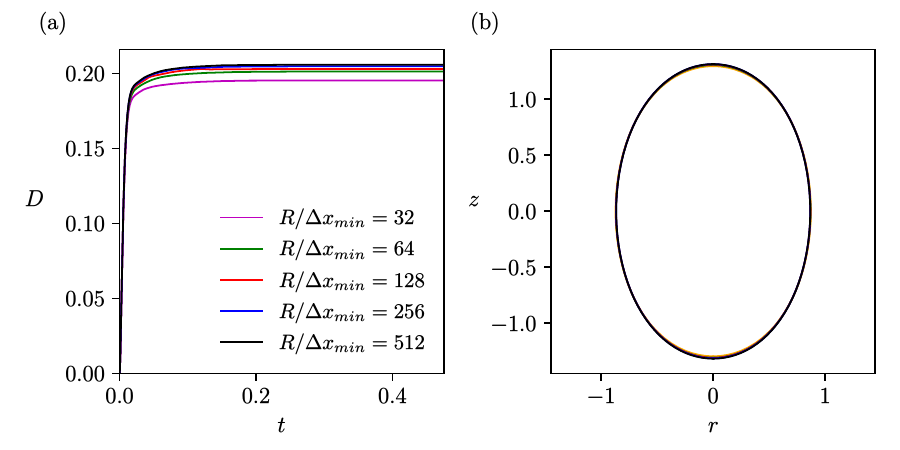}
	\caption{(a) Deformation parameter variation with time at various refinements of the grid. (b) Steady state deformed interface of the drop for various grid refinements. 
    Simulation parameters: $Re=1$, $Ca_E=0.4$, $De=5$, $\sigma_r=10$, $\epsilon_r=1.37$, $\beta=1/9$, $\varepsilon=0.25$}
	\label{Fig_grid_prolate}
\end{figure}
Further, a grid independence study is performed to ensure that the simulation results are not dependent on the grid resolution. We use the adaptive mesh refinement for the numerical simulations and grid refinement is characterized by the parameter $R/\Delta x_{min}$. \autoref{Fig_grid_prolate} shows the grid convergence study. \autoref{Fig_grid_prolate}(a) shows the variation of deformation parameter with time for different grid refinements.  We can see that results for $R/\Delta x_{min}=512$ and $R/\Delta x_{min}=256$ are almost overlapping with each other. \autoref{Fig_grid_prolate}(b) shows the steady state interface shape for various grid refinements. 
The values of deformation parameter and the relative percentage error for various grid refinements are shown in \autoref{Table_gridConvergence}.
As relative percentage error for $R/\Delta x_{min}=256$ is less than 1 percent, we have considered $R/\Delta x_{min}=256$ for conducting the numerical simulations in this study. 
\begin{table}[ht!]
    \centering\caption{The deformation parameter and relative percentage error for different grid refinements}
    \vspace{0.5em}
    \fontsize{11}{20}\selectfont
    \begin{tabular}{c|c|c}
        $R/\Delta x_{min}$ & $D$        & $e_{rel} = \frac{|D - D_{512}|}{|D_{512}|} \times 100$ \\ \hline
        32                 & 0.19522819	& 5.092270099       \\
        64                 & 0.20115761	& 2.209757118       \\
        128                & 0.20295425	& 1.336343172       \\
        256                & 0.20471908	& 0.478393257       \\
        512                & 0.20570315	& 0                 
    \end{tabular}
    \label{Table_gridConvergence}
\end{table}

\section{Results and discussion}\label{Sec_results}
\begin{figure}[ht!]
    \centering
    \includegraphics[width=0.9\textwidth]{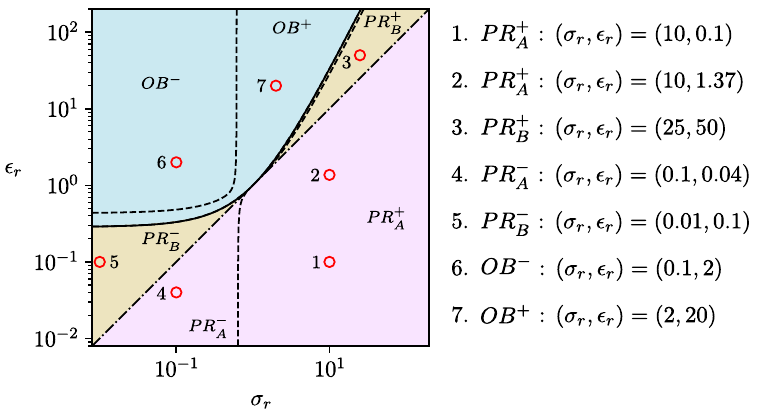}
    \caption{$(\sigma_r, \epsilon_r)$ pairs selected for the study are enumerated and marked (red circle markers) on $(\sigma_r, \epsilon_r)$ phase plot on log-log scale.}
    \label{Fig_CasesSelection}
\end{figure}
The drop deformation is characterized using Taylor’s deformation parameter,
\begin{equation}
    D = \frac{L-B}{L+B}
\end{equation}
where $L$ and $B$ denote the drop dimensions along and perpendicular to the applied electric field, respectively. 

In this study, the electrohydrodynamic deformation of a viscoelastic drop is investigated for $\rho_r=1$, $\mu_r=1$, $Re=1$, $\varepsilon=0.25$, and $\beta_i=1/9$. Six distinct regions on the $(\sigma_r, \epsilon_r)$ plane, denoted $PR_A^+$, $PR_B^+$, $PR_A^-$, $PR_B^-$, $OB^-$, and $OB^+$, identified from asymptotic analysis (\cite{bangar2026large}), serve as the basis for the selection of conductivity ratio and permittivity ratio for the study. Two points from $PR_A^+$ and a representative point from each of the other regions, are selected for the study. Selected points are enumerated and marked on $(\sigma_r, \epsilon_r)$ plot, are shown in \autoref{Fig_CasesSelection}.
For each selected pair, numerical simulations are performed over a range of $Ca_E$ and $De$. The dimensional parameters are fixed as: drop radius $R=0.001$, ambient density $\rho_e = 1000$, ambient permittivity $\epsilon_e = 4.425 \times 10^{-11}$, ambient conductivity $\sigma_e = 10^{-5}$, and interfacial tension $\gamma = 0.065$. Remaining dimensional parameters are obtained from the specified non-dimensional values. The velocity scale, obtained from the balance of electric and viscous stresses, is $U = 0.2549 \sqrt{Ca_E}$ leading to an electric Reynolds number $Re_E = \frac{\epsilon_e U}{R\sigma_e} = 1.128 \times 10^{-3}\sqrt{Ca_E}$.

\begin{figure}
    \centering
    \includegraphics[width=1\linewidth]{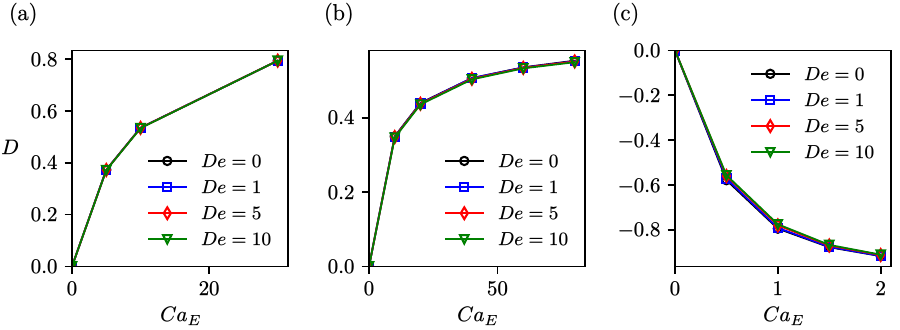}
    \caption{Deformation versus $Ca_E$ for various $De$, for $\rho_r=1$, $\mu_r=1$, $Re=1$, $\beta_i = 1/9$, $\varepsilon=0.25$. (a) $\sigma_r=0.1$, $\epsilon_r=0.04$ $(PR_A^-)$ (b) $\sigma_r=0.01$, $\epsilon_r=0.1$ $(PR_B^-)$ (c) $\sigma_r=2$, $\epsilon_r=20$ $(OB^+)$}
    \label{Fig_neg_D1D2}
\end{figure}

Building on the Newtonian drop analysis of \citet{lac2007axisymmetric} ($De=0$) and the analysis of Oldroyd B drop by \cite{bangar2026large}, the present work extends the investigation to linear PTT drops subjected to steady electric fields. In the $PR_A^-$, $PR_B^-$, and $OB^+$ regions, the first and second order deformation coefficients have opposite signs, and stable drop shapes are attained even at large $Ca_E$. For these regions, \autoref{Fig_neg_D1D2} shows that the deviation from Newtonian behavior is negligible. Thus, for subsequent discussions, we focus on $PR_A^+$, $PR_B^+$ and $OB^-$ regions of $(\sigma_r, \epsilon_r)$ plane.

\subsection{The drop behavior for \texorpdfstring{$(\sigma_r, \epsilon_r)$ from $PR_A^+$}{Lg}}
\subsubsection{The drop behavior for \texorpdfstring{$(\sigma_r, \epsilon_r) = (10, 0.1)$}{Lg}}
\begin{figure}[ht!]
    \centering
    \includegraphics[width=1\textwidth]{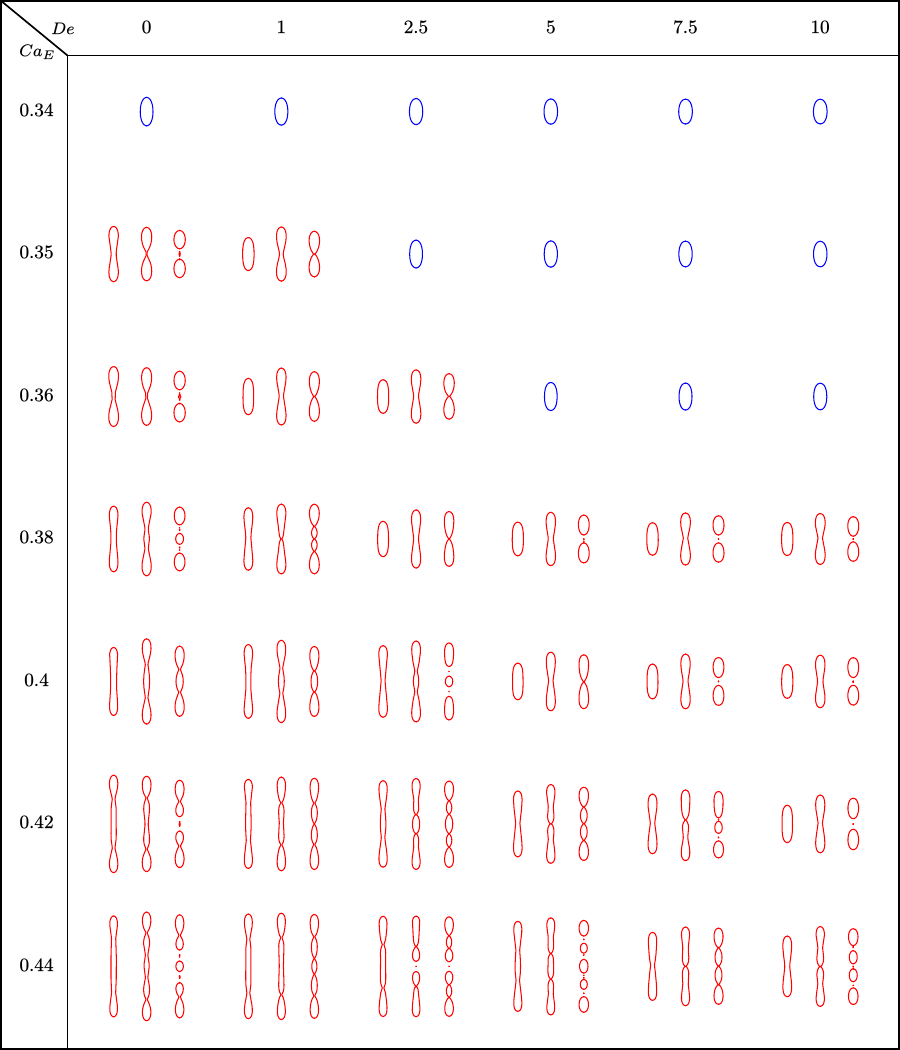}
    \caption{Behavior of drop interface in presence of electric field for various values of $Ca_E$ and $De$ for $(\sigma_r, \epsilon_r) = (10, 0.1)$, $\mu_r=1$, $\rho_r=1$, $Re=1$, $\varepsilon=0.25$, $\beta_i=1/9$}
    \label{Fig_interface_PRA_plus_a}
\end{figure}
\begin{figure}[ht!]
    \centering
    \includegraphics[width=1\textwidth]{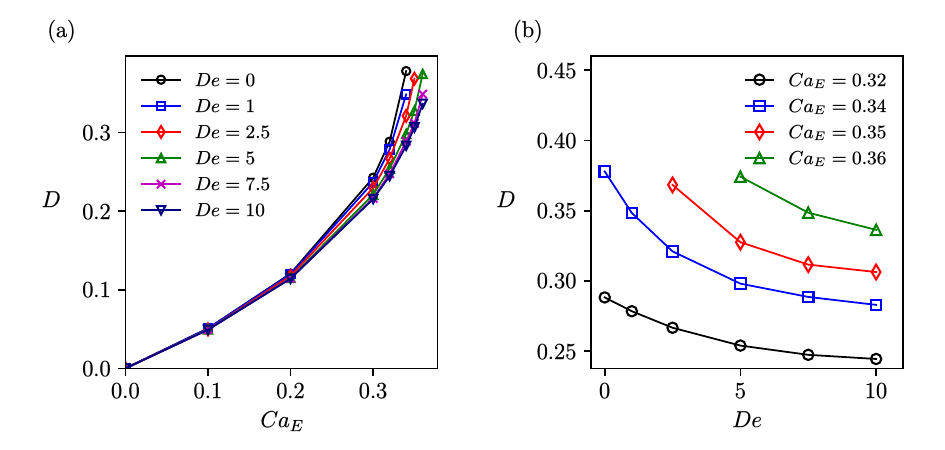}
    \caption{(a) Deformation versus $Ca_E$ for various $De$, (b) Deformation versus $De$ for various $Ca_E$, for $\rho_r=1$, $\mu_r=1$, $Re=1$, $\sigma_r=10$, $\epsilon_r=0.1$, $\beta_i = 1/9$, $\varepsilon = 0.25$. (from $PR_A^+$)}
    \label{Fig_deformation_PRA_plus_a}
\end{figure}
Similar to the Newtonian case, a viscoelastic drop with $(\sigma_r, \epsilon_r) = (10, 0.1)$ in the $PR_A^+$ regime deforms into a stable prolate spheroid under the action of a uniform, steady electric field, provided the electric capillary number ($Ca_E$) remains below a critical threshold. Beyond this threshold, the drop undergoes breakup.
\autoref{Fig_interface_PRA_plus_a} presents the time evolution of the drop interface for various combinations of $Ca_E$ and Deborah number ($De$). For cases resulting in a stable spheroidal deformation, only the final equilibrium interface is plotted (blue). When a stable configuration is not attained, the interface at two intermediate times and at the terminal simulation time is shown (red). These snapshots are chosen to highlight the characteristic stages of deformation and breakup for each $(Ca_E,De)$ pair.
The figure reveals a distinct dependence of the breakup threshold on viscoelasticity. At higher Deborah numbers ($De \geq 5$), the drop retains a stable spheroidal shape for $Ca_E \leq 0.36$, but transitions to multi-lobed deformations and eventual breakup once $Ca_E \geq 0.37$, placing the critical capillary number $Ca_E^{crit}$ in the range $(0.36,,0.37)$. In contrast, at lower Deborah numbers ($De \leq 1$), spheroidal stability persists only up to $Ca_E \leq 0.34$, while instability and breakup occur for $Ca_E \geq 0.35$, indicating $Ca_E^{crit} \in (0.34,,0.35)$.
These results clearly demonstrate that the critical electric capillary number increases with increasing Deborah number, underscoring the stabilizing influence of viscoelasticity on drop deformation under electric fields.

\autoref{Fig_deformation_PRA_plus_a}(a) illustrates the variation of stable spheroidal drop deformation with the electric capillary number ($Ca_E$) for $(\sigma_r, \epsilon_r) = (10, 0.1)$ at different Deborah numbers ($De$). Similar to Newtonian drops, the deformation curves exhibit a positive curvature across all $De$. However, for a fixed $Ca_E$, the extent of deformation decreases as $De$ increases, highlighting the elastic resistance of the drop to deformation.
This trend is further quantified in \autoref{Fig_deformation_PRA_plus_a}(b), which depicts the dependence of deformation on $De$ for various $Ca_E$. The results show a monotonic reduction in deformation with increasing $De$, followed by a saturation at higher Deborah numbers, indicating that beyond a certain elasticity, further increases in $De$ exert a diminishing influence on drop deformation.

\begin{figure}[ht!]
    \centering
    \includegraphics[width=0.99\textwidth]{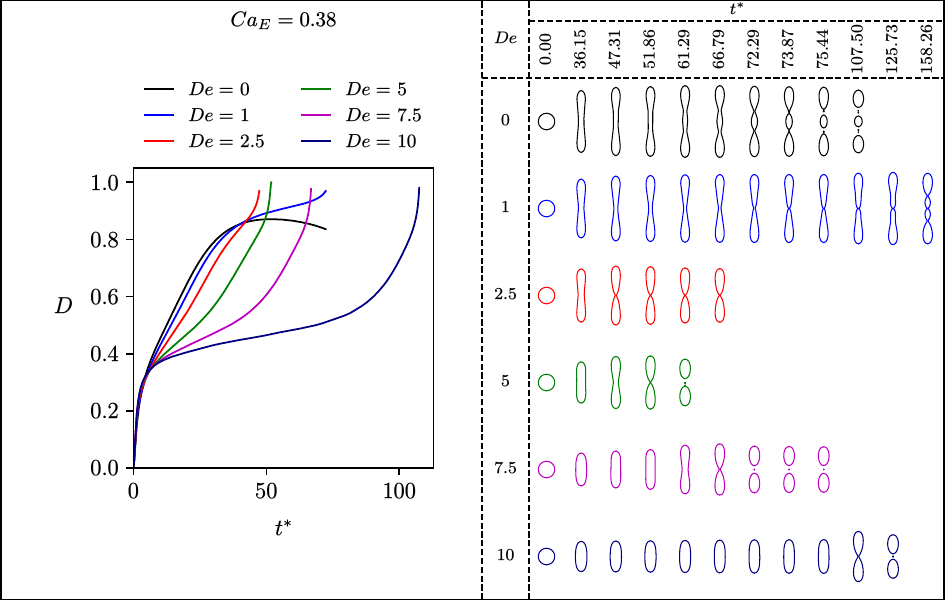}
    \caption{Evolution of drop interface with time for  $\rho_r=1$, $\mu_r=1$, $Re=1$, $\sigma_r=10$, $\epsilon_r=0.1$, $\beta_i = 1/9$, $\varepsilon = 0.25$, $Ca_E=0.38$. Deformation versus time is shown for each $De$ till first pinch off.}
    \label{Fig_timeEvolution_PRA_plus_a_Cap38}
\end{figure}
\begin{figure}[ht!]
    \centering
    \includegraphics[width=1\textwidth]{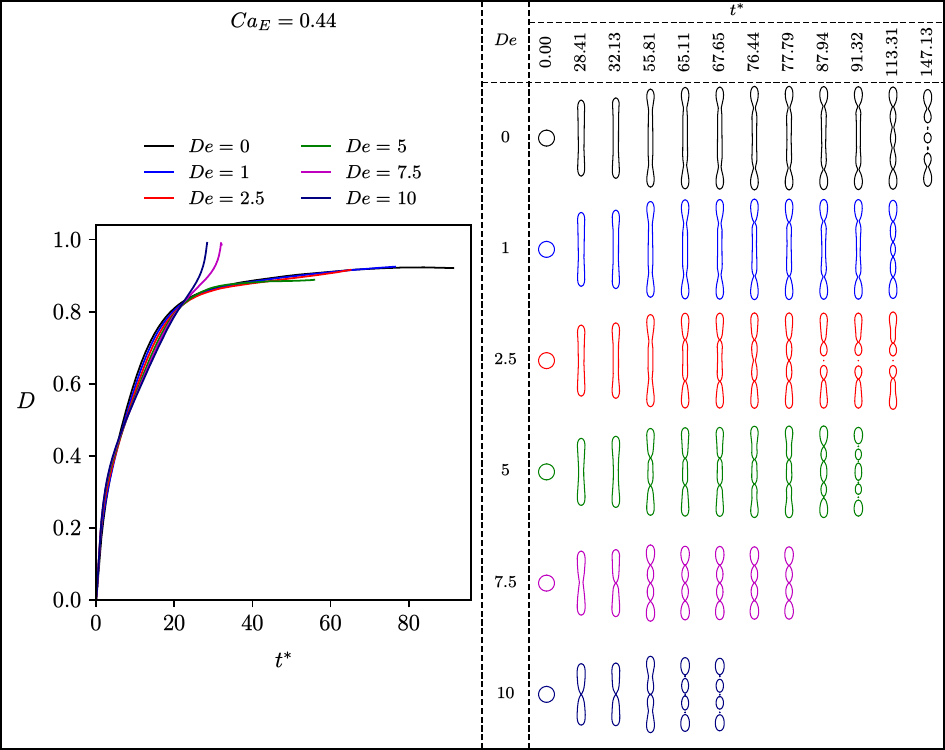}
    \caption{Evolution of drop interface with time for  $\rho_r=1$, $\mu_r=1$, $Re=1$, $\sigma_r=10$, $\epsilon_r=0.1$, $\beta_i = 1/9$, $\varepsilon = 0.25$, $Ca_E=0.44$. Deformation versus time is shown for each $De$ till first pinch off.}
    \label{Fig_timeEvolution_PRA_plus_a_Cap44}
\end{figure}
\autoref{Fig_timeEvolution_PRA_plus_a_Cap38} illustrates the temporal evolution of the drop interface at a constant electric capillary number ($Ca_E=0.38$) across various Deborah numbers ($De$). 
At $Ca_E=0.38$, the breakup morphology varies significantly with the Deborah number ($De$). For $De \geq 2.5$, the drop primarily undergoes a two-lobed breakup. Specifically, for $De=5,7.5$, and 10, these lobes initially separate but then approach each other, while at $De=2.5$, the two lobes remain attached (or immediately reattach upon breakup). When $De=0$, a three-lobed breakup is observed. In contrast, at $De=1$, an initial two-lobed breakup occurs, with each resulting lobe subsequently undergoing further breakup to yield a four-lobed structure.
The temporal evolution of deformation parameter for various $De$ values, up to the first breakup event, is also illustrated. A rapid increase in deformation precedes breakup, particularly for cases exhibiting two-lobed initial breakup. We note a trend, where, the initial number of breakup lobes decreases with increasing $De$: three lobes for $De=0$, and two lobes for $De 
\geq 1$. As previously mentioned, for $De=1$, these elongated initial lobes undergo secondary breakup, leading to a four-lobed final structure. Significantly, for $De \geq 2.5$, the breakup time increases with increasing $De$, indicating a delayed breakup for higher Deborah numbers.

Moving to a higher electric capillary number, $Ca_E=0.44$, \autoref{Fig_timeEvolution_PRA_plus_a_Cap44} depicts the temporal evolution of the drop interface, revealing more complex breakup patterns.
At $Ca_E=0.44$, the breakup behavior varies distinctly across different Deborah numbers ($De$). 
For $De=0$, the drop elongates to form a three-lobed structure with a longer central lobe, which further breaks into three lobes, finally resulting in a central lobe with two-lobed structures at both ends.
For $De=1$, a three-lobed structure with a longer central lobe forms; the end lobes pinch off, and the central lobe subsequently breaks into three, leading to five attached lobes.
At $De=2.5$, the drop forms a four-lobed structure where the end lobes pinch off first, followed by the central two-lobed structure breaking into two attached two-lobed entities.
When $De=5$, the initial breakup is three-lobed and attached, where the end lobes then split further into two, resulting in five separated lobes.
Finally, for $De=7.5$ and 10, an initial two-lobed breakup occurs; these lobes remain attached, with each subsequently breaking into two, yielding a total of four lobes (attached at $De=7.5$ but separated at $De=10$). 

Similar to the previous case, a sharp increase in deformation precedes the initial breakup for $De=7.5$ and $De=10$, because of two lobed initial breakup. 
Overall, the initial number of breakup lobes decreases with increasing $De$ (two for $De=7.5,10$ and three for $De \leq 5$). For initial three-lobed breakups, the central lobe length decreases with increasing $De$. This results in a three-lobed breakup of the central lobe at lower $De$ (0,1), a two-lobed breakup at slightly higher $De=2.5$, and breakup of the end lobes at $De=5$ (where the central lobe is smaller than the end lobes). Interestingly, for the initial two-lobed breakups ($De=7.5,10$), the drop breaks up earlier at higher $De=10$ than at $De=7.5$, a contrast to the behavior observed at $Ca_E=0.38$.

In general, an increase in $De$ enhances the viscoelastic drop's resistance to prolate deformation. This resistance manifests differently depending on $Ca_E$.
At $Ca_E=0.38$, the breakup time increases with increasing $De$. Initially, electric forces cause rapid deformation, followed by a slower increase due to viscoelastic resistance at intermediate times. However, once a critical deformation threshold is crossed, the capillary forces rapidly increase the drop deformation, leading to the drop breakup. The duration of this slower deformation phase is longer at higher $De$, resulting in delayed breakup.
In contrast, at $Ca_E=0.44$, higher electric forces cause the drop to reach larger deformations more rapidly during the initial phase. Beyond this, higher $De$ values lead to quicker breakup.
This contrasting behavior implies that viscoelasticity initially impedes deformation but can accelerate breakup once a critical deformation threshold is exceeded, explaining the different breakup dynamics observed at varying $Ca_E$. Furthermore, the lower influence of viscoelasticity at lower $De$ allows for a longer initial rapid deformation phase, contributing to the formation of a greater number of lobes.

\subsubsection{The drop behavior for \texorpdfstring{$(\sigma_r, \epsilon_r) = (10, 1.37)$}{Lg}}
\begin{figure}[ht!]
    \centering
    \includegraphics[width=1\textwidth]{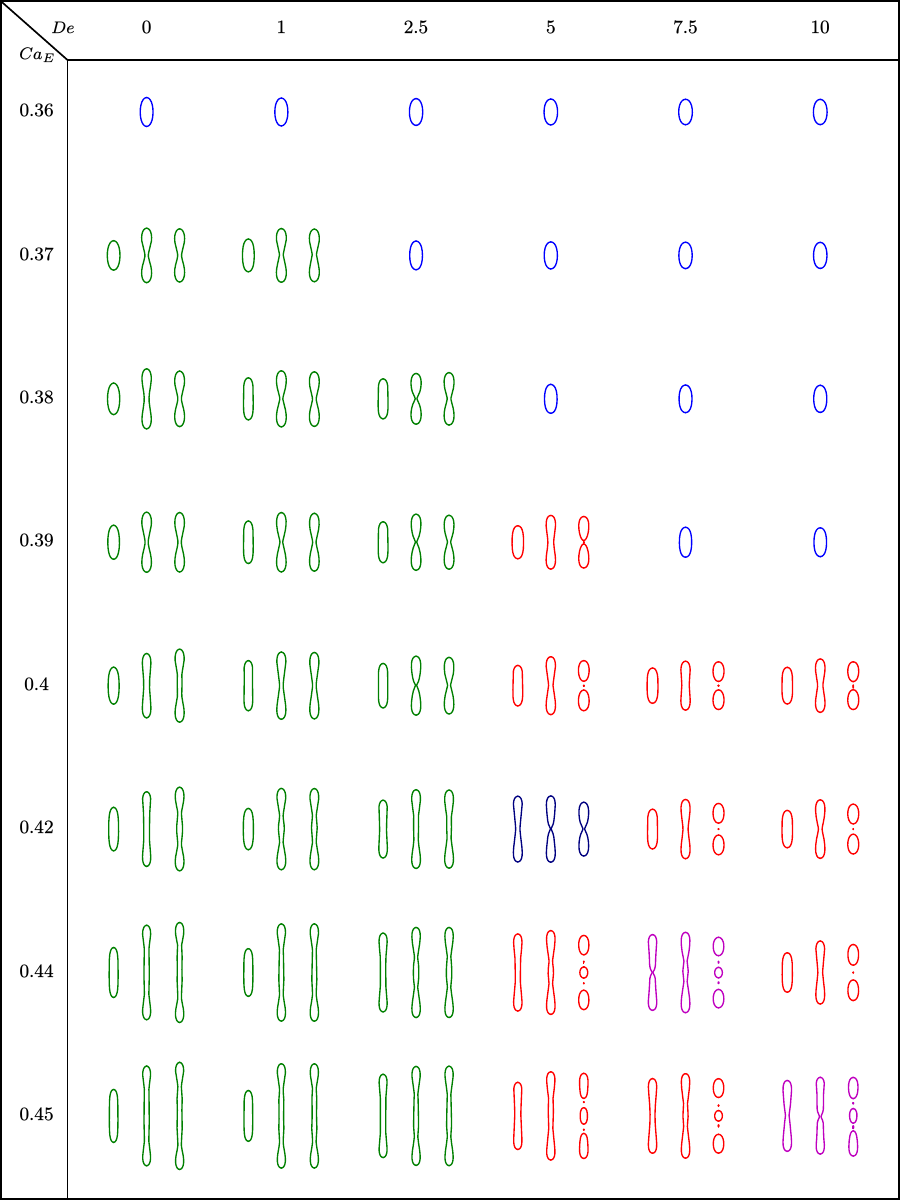}
    \caption{Behavior of drop interface in presence of electric field for various values of $Ca_E$ and $De$ for $(\sigma_r, \epsilon_r) = (10, 1.37)$, $\mu_r=1$, $\rho_r=1$, $Re=1$, $\varepsilon=0.25$, $\beta_i=1/9$}
    \label{Fig_interface_PRA_plus_b}
\end{figure}
\autoref{Fig_interface_PRA_plus_b} illustrates the temporal evolution of a viscoelastic drop interface for $(\sigma_r, \epsilon_r) = (10, 1.37)$ across various electric capillary numbers ($Ca_E$) and Deborah numbers ($De$). Distinct color codes indicate different behaviors within the figure. A blue interface signifies a stable spheroidal equilibrium shape. For cases where the drop achieves stable multi-lobed configurations, the interface is shown in green at two intermediate time instances and at its final steady-state. When the drop undergoes breakup, the interface is depicted in red at two intermediate time instances and at the terminal simulation time. It is particularly interesting to note the cases shown in magenta: here, the drop initially deforms into a two-lobed shape and appears on the verge of breakup, but its neck recovers to form a three-lobed structure, which subsequently breaks into three daughter drops. Additionally, for the case highlighted in navy blue, the drop deforms into a two-lobed structure seemingly poised for pinch-off, yet it recovers to form a stable two-lobed configuration.

\begin{figure}
    \centering
    \includegraphics[width=1\textwidth]{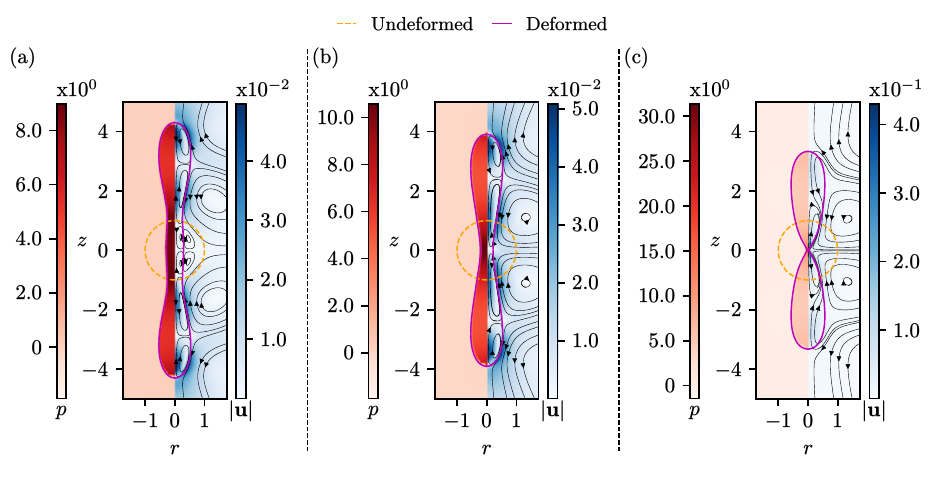}
    \caption{Streamlines and pressure contours for $Ca_E=0.4$, $(\sigma_r, \epsilon_r) = (10, 1.37)$, $\mu_r=1$, $\rho_r=1$, $Re=1$, $\varepsilon=0.25$, $\beta_i=1/9$. (a) $De=0$ (b) $De=1$ (c) $De=5$}
    \label{Fig_streamlines_pressure}
\end{figure}

The stability of multi-lobed drop shapes is analyzed by examining streamlines and pressure contours, as illustrated in \autoref{Fig_streamlines_pressure}. The streamlines and  pressure contours are presented for a constant capillary number ($Ca_E = 0.4$) across three distinct Deborah numbers ($De$).
For $De = 0$ and $De = 1$, which correspond to stable three-lobed and two-lobed drop configurations, respectively, the streamlines are observed to be tangential to the drop interface. The flow thus indicates a lack of interface motion, which is characteristic of a stable drop shape. Concurrently, organized internal flow patterns are observed, six recirculating eddies are present for $De = 0$, and four for $De = 1$. These coherent flow structures within the drop counteract the destabilizing capillary forces and prevent drop breakup.
In contrast, at $De = 5$, the drop undergoes breakup. The streamlines and pressure field, for the time instance before the breakup, reveal a significant deviation from the stable state. In this unstable regime, the streamlines are no longer tangential to the interface, signifying interfacial motion. A high-magnitude outward velocity in the neck region of the deforming drop leads to the drop breakup.

\begin{figure}[ht!]
    \centering
    \includegraphics[width=1\textwidth]{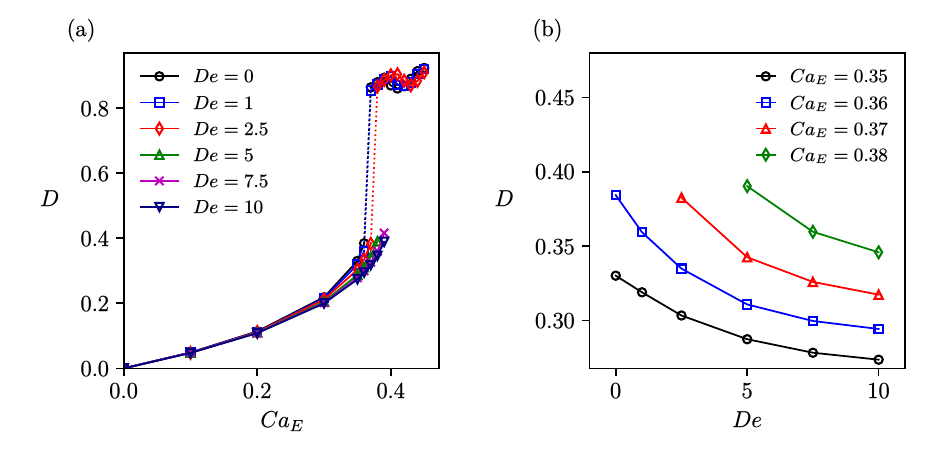}
    \caption{(a) Deformation versus $Ca_E$ for various $De$, (b) Deformation versus $De$ for various $Ca_E$, for $\rho_r=1$, $\mu_r=1$, $Re=1$, $\sigma_r=10$, $\epsilon_r=1.37$, $\beta_i = 1/9$, $\varepsilon = 0.25$. (from $PR_A^+$)}
    \label{Fig_deformation_PRA_plus_b}
\end{figure}
\begin{figure}[ht!]
    \centering
    \includegraphics[width=1\textwidth]{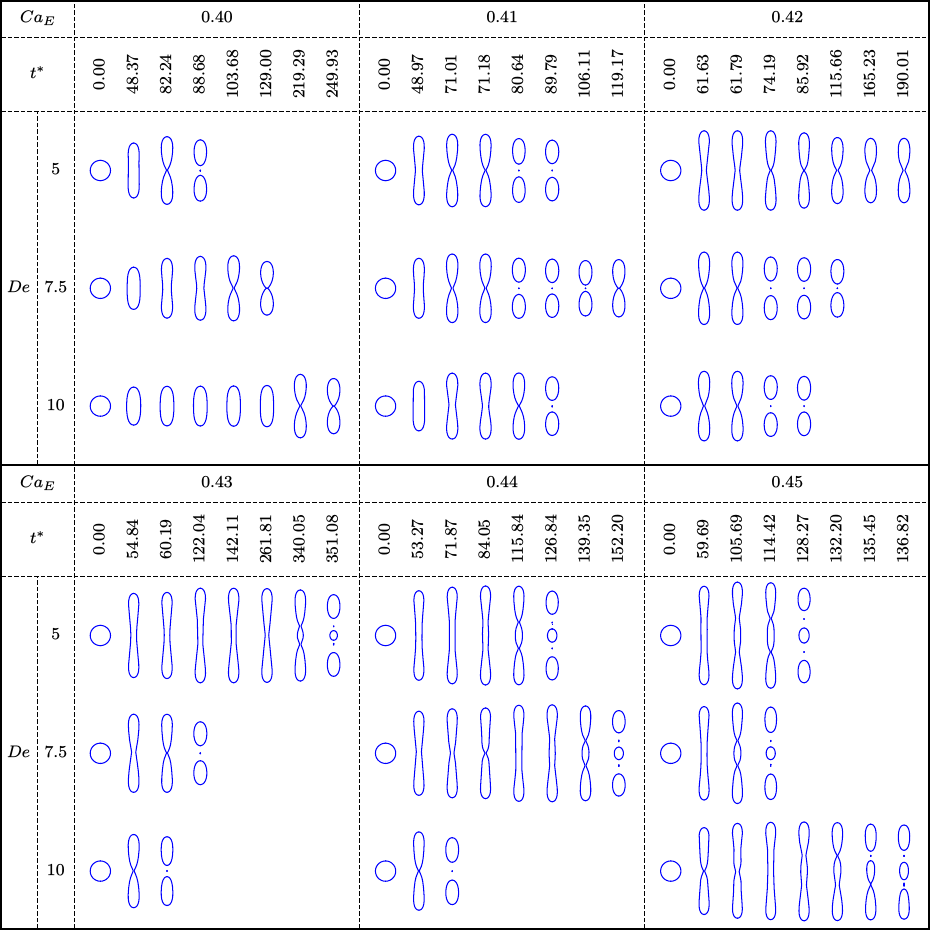}
    \caption{Temporal evolution of drop interface for $De=5,7.5,10$ across the range of $Ca_E$ for $\rho_r=1$, $\mu_r=1$, $Re=1$, $\sigma_r=10$, $\epsilon_r=1.37$, $\beta_i = 1/9$, $\varepsilon = 0.25$.}
    \label{Fig_timeEvolution_PRA_plus_b}
\end{figure}

\autoref{Fig_deformation_PRA_plus_b}(a) presents the relationship between drop deformation and the electric capillary number for $(\sigma_r, \epsilon_r) = (10, 1.37)$ across various Deborah numbers. We observe two distinct behaviors based on the Deborah number range. For lower Deborah numbers ($De=0, 1, 2.5$), the deformation curve exhibits two distinct branches. The lower branch corresponds to stable spheroidal deformation, up to a certain critical $Ca_E$. Beyond this critical value, the drop deforms into stable multi-lobed shapes, which are represented by the upper branch of the curve. In contrast, for higher Deborah numbers ($De=5,7.5,10$), only a single branch of the deformation curve is observed. In these scenarios, the drop maintains a stable spheroidal shape up to a critical $Ca_E$, and upon exceeding this value, it undergoes breakup, thereby precluding the formation of stable multi-lobed shapes.
Across all Deborah numbers, the lower branch of the deformation curve consistently exhibits a positive curvature. We also note that for a given $Ca_E$, the drop deformation is lower at higher Deborah numbers, indicating a general decrease in drop deformation as $De$ increases. This trend is further substantiated and clearly visible in \autoref{Fig_deformation_PRA_plus_b}(b), which illustrates the direct dependence of deformation on the Deborah number. Moreover, \autoref{Fig_deformation_PRA_plus_b}(b) also reveals that the drop deformation tends to saturate at higher values of the Deborah number.

\autoref{Fig_timeEvolution_PRA_plus_b} presents the temporal evolution of the viscoelastic drop interface for multi-lobed breakup regimes ($De\geq 5$), as a function of the electric capillary number ($Ca_E$).
At a lower $Ca_E ~(=0.4)$, the breakup time is observed to increase with $De$. However, at a slightly elevated $Ca_E=0.41$, the overall breakup time decreases across all Deborah numbers examined. Specifically, $De=5$ and $De=7.5$ exhibit similar breakup times, while $De=10$ shows a marginally longer breakup time compared to $De=5$ and $De=7.5$. Notably, at $Ca_E=0.42$, for $De=5$, the drop initially approaches a two-lobed breakup but then recovers to attain a stable two-lobed steady shape. For $Ca_E=0.42$, drop exhibits two-lobed breakup for both $De=7.5$ and 10, with $De=10$ showing a shorter breakup time than $De=7.5$. In contrast, $De=5$ undergoes a three-lobed breakup, which occurs at a longer breakup time.
For $Ca_E=0.44$, $De=10$ still shows rapid two-lobed breakup, whereas $De=7.5$ recovers from a two lobed configuration on the verge of breakup, to undergo a three-lobed breakup with a longer breakup time than the three-lobed breakup of $De=5$. At $Ca_E=0.45$, the time taken for the three-lobed breakup decreases for both $De=5$ and 7.5, with a more pronounced reduction for $De=7.5$, leading to a shorter breakup time compared to $De=5$; and $De=10$ transitions from a near two-lobed breakup to three-lobed shape and subsequently breaks into three daughter droplets.
In summary, these observations reveal a nuanced influence of Deborah number on breakup time depending on the electric capillary number. Near the critical $Ca_E$ for two-lobed breakup, higher $De$ tends to delay breakup. However, as $Ca_E$ increases beyond this critical value, the breakup time decreases more rapidly for higher values of $De$, suggesting an accelerating effect. A similar pattern is observed for three-lobed breakup: higher $De$ delays breakup near the critical $Ca_E$ for this mode, but accelerates it at higher $Ca_E$ values.

\subsection{The drop behavior for \texorpdfstring{$(\sigma_r, \epsilon_r)$ from $PR_B^+$}{Lg}}
\begin{figure}[ht!]
    \centering
    \includegraphics{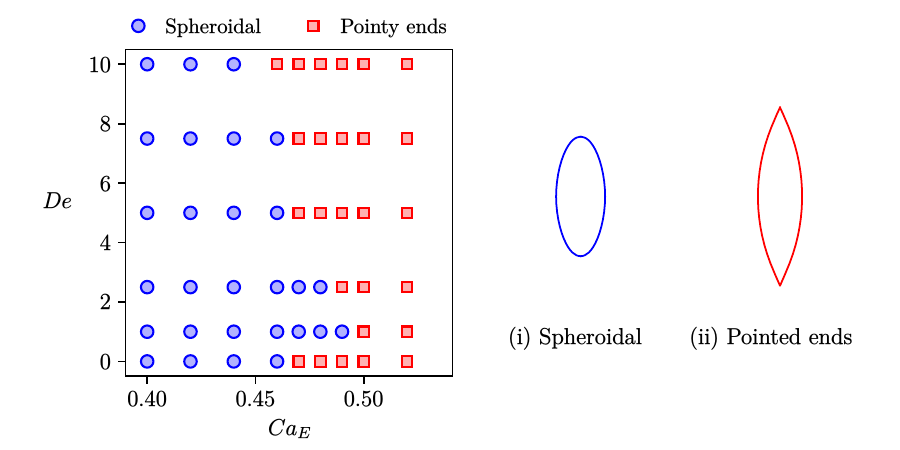}
    \caption{Phase plot on $(Ca_E, De)$ plane representing the existence of spheroidal shapes and shapes with pointed ends}
    \label{Fig_phase_plot_PRB_plus}
\end{figure}
Similar to a Newtonian drop, a viscoelastic drop with $(\sigma_r, \epsilon_r)$ from the $PR_B^+$ region, when subjected to an electric field, deforms into a spheroidal shape up to a critical electric capillary number ($Ca_E^{crit}$). Beyond this threshold, the drop forms characteristic pointed-end shapes. \autoref{Fig_phase_plot_PRB_plus} illustrates this behavior as a phase diagram in the $(Ca_E, De)$ plane. The diagram delineates regions of stable spheroidal shapes, marked by blue, from those forming pointed-end shapes (non-dimensional curvature of order 100), indicated by red markers. For Deborah numbers $De \geq 1$, the diagram distinctly shows that the maximum electric capillary number permitting a stable spheroidal shape decreases with increasing Deborah number.

\begin{figure}[ht!]
    \centering
    \includegraphics[width=1\textwidth]{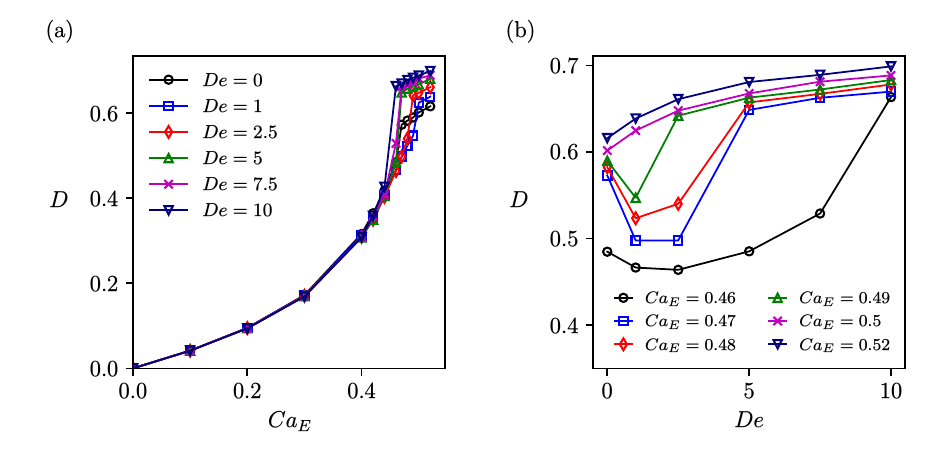}
    \caption{(a) Deformation versus $Ca_E$ for various $De$, (b) Deformation versus $De$ for various $Ca_E$, for $\rho_r=1$, $\mu_r=1$, $Re=1$, $\sigma_r=25$, $\epsilon_r=50$, $\beta_i = 1/9$, $\varepsilon = 0.25$. (from $PR_B^+$)}
    \label{Fig_deformation_PRB_plus}
\end{figure}
\begin{figure}[ht!]
    \centering
    \includegraphics[width=1\textwidth]{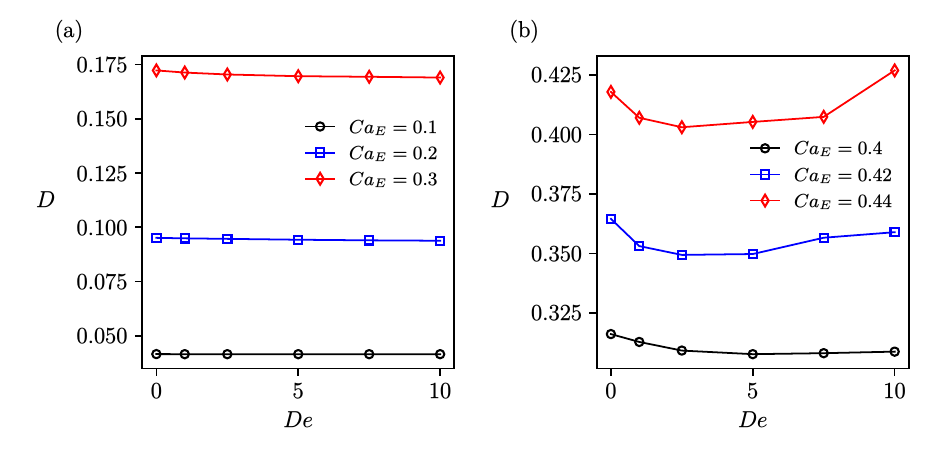}
    \caption{(a) Deformation versus $De$ for various $Ca_E$ for lower electric capillary numbers, (b) Deformation versus $De$ for various $Ca_E$ for intermediate electric capillary numbers, for $\rho_r=1$, $\mu_r=1$, $Re=1$, $\sigma_r=25$, $\epsilon_r=50$, $\beta_i = 1/9$, $\varepsilon = 0.25$. (from $PR_B^+$)}
    \label{Fig_deformation_PRB_plus_2}
\end{figure}
\begin{figure}[ht!]
    \centering
    \includegraphics[width=1\textwidth]{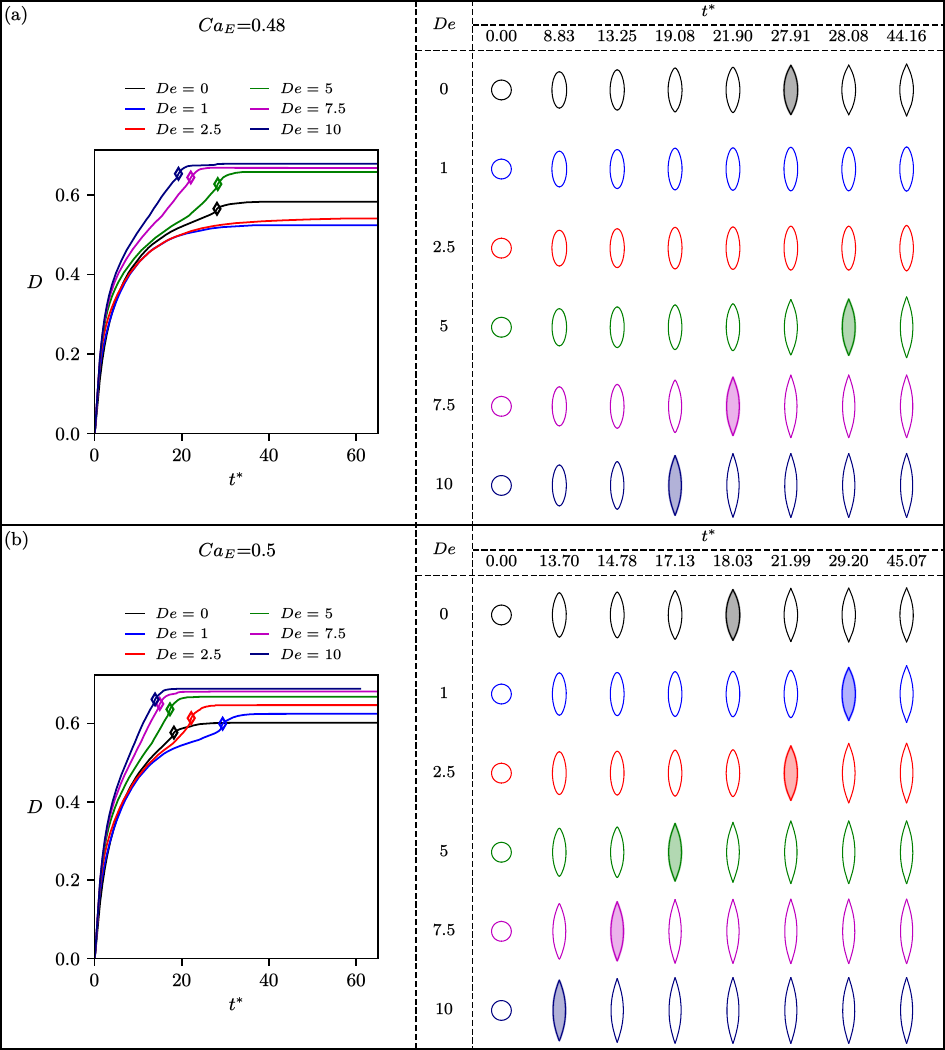}
    \caption{Variation of deformation parameter with time and temporal evolution of drop interface for various $De$ for $\rho_r=1$, $\mu_r=1$, $Re=1$, $\sigma_r=25$, $\epsilon_r=50$, $\beta_i = 1/9$, $\varepsilon = 0.25$. (a) $Ca_E=0.48$ (b) $Ca_E=0.5$. Markers on each $D$ versus $t^*$ indicates the time at which drop transforms to the shape characterized by the pointy ends. Interface shape corresponding to the diamond marker is filled with the lighter shade of the corresponding color.}
    \label{Fig_timeEvolution_PRB_plus}
\end{figure}
\begin{figure}[ht!]
    \centering
    \includegraphics[width=1\textwidth]{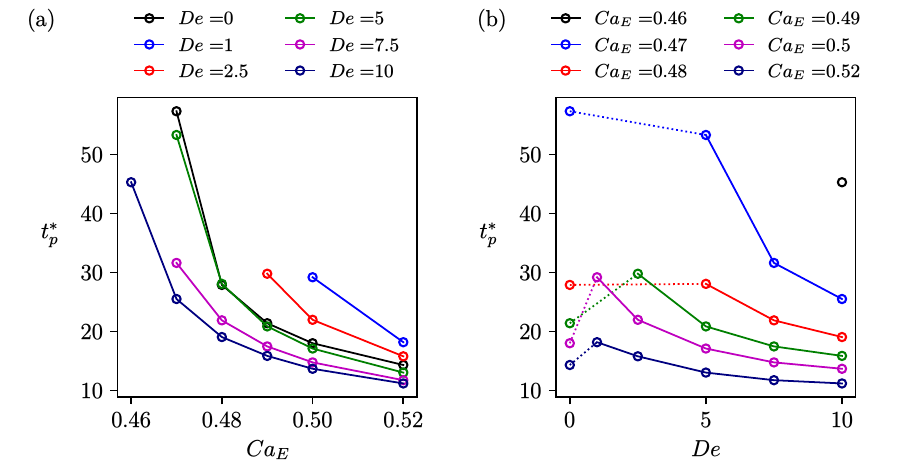}
    \caption{(a) The variation of time that is required for the formation of pointed ends with $Ca_E$. (b) Time of formation of pointed ends versus $De$ for $\rho_r=1$, $\mu_r=1$, $Re=1$, $\sigma_r=25$, $\epsilon_r=50$, $\beta_i = 1/9$, $\varepsilon = 0.25$.}
    \label{Fig_pointyEndsTime}
\end{figure}
The influence of both the electric capillary number ($Ca_E$) and the Deborah number ($De$) on drop deformation is comprehensively illustrated in \autoref{Fig_deformation_PRB_plus} and \ref{Fig_deformation_PRB_plus_2}. 
\autoref{Fig_deformation_PRB_plus}(a) presents the variation of drop deformation with $Ca_E$ across different $De$. For all Deborah numbers, deformation consistently increases with $Ca_E$. The concave upward nature of these deformation curves further indicates an accelerated increase in deformation at higher $Ca_E$. A notable feature observed for every $De$ is a sudden jump in deformation at a certain critical $Ca_E$. This abrupt increase signifies the transition from a stable spheroidal shape to a more elongated configuration with pointed ends, after which deformation continues to increase at a nearly constant rate with $Ca_E$.
\autoref{Fig_deformation_PRB_plus}(b) illustrates drop deformation variation with $De$ for several higher $Ca_E$ values, specifically those where the drop exhibits pointed ends for at least one $De$. For $Ca_E \geq 0.5$, where the drop consistently deforms into shapes with pointed ends across all $De$, deformation is observed to increase gradually with $De$. The behavior is non-monotonic at $Ca_E=0.46$: deformation initially decreases, then increases with increasing $De$. It is crucial to note that at $Ca_E=0.46$, the drop forms pointed ends exclusively at $De=10$, resulting in the highest deformation for $De=10$. For intermediate $Ca_E$ values ($0.47,0.48,0.49$), pointed ends appear for $De=0$ and also at higher $De$, while for some intermediate $De$ values drop maintains stable spheroidal shape.
Further insights into the dependence of drop deformation on $De$ for low to moderate $Ca_E$ are provided in \autoref{Fig_deformation_PRB_plus_2}. 
Specifically, \autoref{Fig_deformation_PRB_plus_2}(a) illustrates the deformation vs. $De$ at very low $Ca_E$, revealing a minimal sensitivity of drop deformation to variations in De, characterized by a slight decrease in deformation with $De$.
In contrast, \autoref{Fig_deformation_PRB_plus_2}(b) depicts the drop deformation as a function of $De$ for moderate $Ca_E$. This regime exhibits a non-monotonic relationship, where droplet deformation initially decreases with increasing $De$, followed by a subsequent increase.

\autoref{Fig_timeEvolution_PRB_plus} presents the temporal evolution of droplet deformation and the corresponding interfacial dynamics for varying Deborah numbers at two distinct electric capillary numbers, $Ca_E=0.48$ and 0.5. Diamond markers on the deformation versus time curves indicate the transition of drop shape from spheroidal to the shape with pointed ends. The shape of the interface corresponding to this time is filled by the lighter shade of the corresponding color. At $Ca_E=0.48$, shapes with pointed ends are observed for $De=0,5,7.5,10$, with $De=1$ and 2.5 resulting in stable spheroidal configurations. For $De \geq 5$ at this $Ca_E$, the transition to a pointed end shape occurs earlier with increasing $De$, while the transition time for $De=0$ is intermediate between $De=5$ and 7.5. At a higher $Ca_E=0.5$, all investigated $De$ values exhibit a transition to the shapes with pointed ends. For $De \geq 1$ at $Ca_E=0.5$, the transition time decreases with increasing $De$, and the transition time for $De=0$ is intermediate between $De=2.5$ and $De=5$.

\autoref{Fig_pointyEndsTime}(a) and (b) illustrate the dependence of the dimensionless transition time to a shape with pointed ends, denoted as $t_p^*$, on the electric Capillary number ($Ca_E$) and the Deborah number ($De$), respectively. From \autoref{Fig_pointyEndsTime}(a), a monotonic decrease in $t_p^*$ with increasing $Ca_E$ is evident across all $De$, signifying an accelerated transition to the shapes with pointed ends at higher electric field strengths. \autoref{Fig_pointyEndsTime}(b) reveals that for $De \geq 1$, $t_p^*$ decreases with increasing $De$, indicating a faster onset of drop shapes with pointed ends at higher Deborah numbers. However, the transition time for $De=0$ is comparable to that observed for $De=5$. This proximity is further corroborated by the near overlap of the $t_p^*$ versus $Ca_E$ curves for $De=0$ and $De=5$ in \autoref{Fig_pointyEndsTime}(a).

\subsection{The drop behavior \texorpdfstring{$(\sigma_r, \epsilon_r)$ from $OB^-$}{Lg}}
\begin{figure}[ht!]
    \centering
    \includegraphics{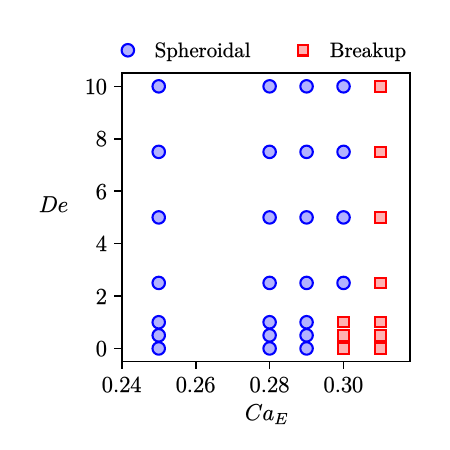}
    \caption{Phase plot on $(Ca_E, De)$ plane representing the existence of stable deformed shapes and drop breakup}
    \label{Fig_phase_plot_OB_minus}
\end{figure}
Similar to its Newtonian counterpart, a viscoelastic drop subjected to an external electric field, for $(\sigma_r, \epsilon_r) = (0.1, 2)$ corresponding to the $OB^-$ region, undergoes oblate deformation up to a critical electric capillary number, and beyond this critical $Ca_E$, the drop loses its stability and experiences electrohydrodynamic breakup.
\autoref{Fig_phase_plot_OB_minus} presents a phase diagram in the $(Ca_E, De)$ plane, delineating the steady-state drop shape and drop breakup regimes. The blue markers denote the existence of stable, steady-state oblate drop shapes, while the red markers denote the occurrence of drop breakup. 
For lower Deborah numbers ($De = 0$, $0.5$, and $1$), drop breakup is observed at an electric capillary number of $Ca_E = 0.3$. In contrast, for higher Deborah numbers ($De = 2.5$, $5$, $7.5$, and $10$), the drop attains a stable oblate shape at $Ca_E = 0.3$, with breakup occurring at a slightly elevated electric capillary number of $Ca_E = 0.31$.

\begin{figure}[ht!]
    \centering
    \includegraphics[width=1\textwidth]{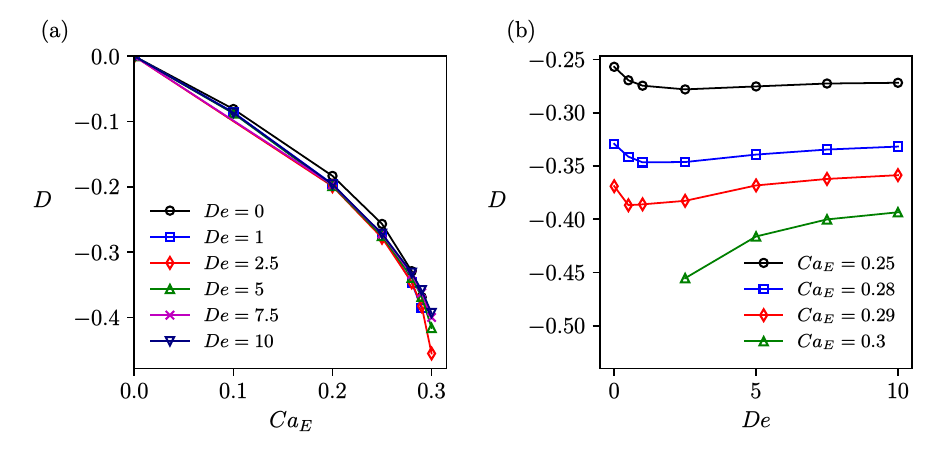}
    \caption{(a) Deformation versus $Ca_E$ for various $De$, (b) Deformation versus $De$ for various $Ca_E$, for $\rho_r=1$, $\mu_r=1$, $Re=1$, $\sigma_r=0.1$, $\epsilon_r=2$, $\beta_i = 1/9$, $\varepsilon = 0.25$. (from $OB^-$)}
    \label{Fig_deformation_OB_minus}
\end{figure}
Figure \ref{Fig_deformation_OB_minus} illustrates the dependence of the drop deformation parameter on the electric capillary number ($Ca_E$) and the Deborah number ($De$).
Specifically, \autoref{Fig_deformation_OB_minus}(a) depicts the evolution of the drop deformation parameter as a function of the electric capillary number for several Deborah numbers. The results indicate that the magnitude of the drop deformation increases with increasing electric capillary number. Consistent with the Newtonian case, the rate of increase in the magnitude of deformation with respect to $Ca_E$ remains positive across the investigated range of Deborah numbers.
\autoref{Fig_deformation_OB_minus}(b) presents the variation of the drop deformation parameter with the Deborah number for various electric capillary numbers. An interesting  non-monotonic behavior of the deformation parameter with the Deborah number is observed. 
An initial increase in the magnitude of drop deformation is observed with an increase in the Deborah number, for a fixed electric capillary number, up to a certain $De$ threshold, beyond which the magnitude of deformation begins to decrease. For example, at $Ca_E = 0.25$, the magnitude of the deformation increases with $De$ until $De = 2.5$, followed by a subsequent decrease. Similarly, at $Ca_E = 0.28$, the magnitude of the deformation increases until $De = 1$ and then decreases. For $Ca_E = 0.29$, the magnitude of the deformation increases with $De$ up to $De = 0.5$ and decreases thereafter. In particular, at $Ca_E = 0.3$, stable droplet shapes are not attainable at lower Deborah numbers ($De = 0$, $0.5$ and $1$), and for higher Deborah numbers where steady states are observed, the magnitude of deformation decreases with increasing $De$. A similar non-monotonic behavior has been observed by \citet{aggarwal2008effects} for the shear-induced deformation of a drop in viscoelastic matrix.

For lower $Ca_E$, increase in magnitude of deformation is observed upto higher values of $De$, before it starts to decrease. For example, for $Ca_E=0.25$, magnitude of deformation increases with $De$ till $De=2.5$ before it starts to decrease. whereas, for $Ca_E=0.29$, magnitude of deformation increases with $De$ till $De=0.5$ and then it shows increase with $De$. Similarly, for $Ca_E=0.28$, magnitude of deformation increases with $De$ till $De=1$ and then it starts to decrease. At $Ca_E=0.3$, drop cannot achieve stable shapes for lower $De$ ($De=0$, 0.5 and 1), and magnitude of deformation decreases with $De$ for higher $De$, where steady shapes are observed.

\begin{figure}
    \centering
    \includegraphics[width=1\textwidth]{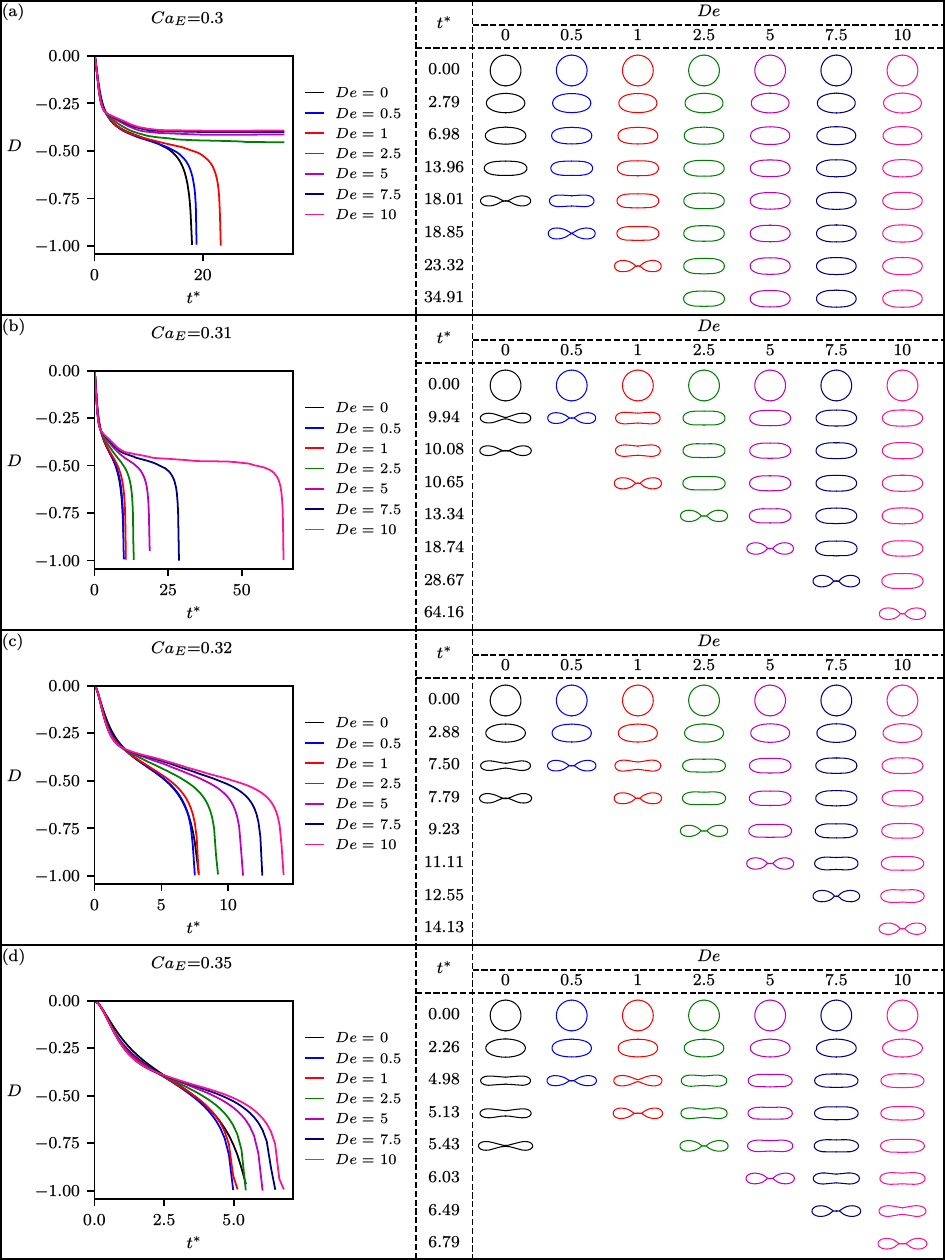}
    \caption{The variation of drop deformation with time and the evolution of the drop interface for various Deborah numbers for $\rho_r=1$, $\mu_r=1$, $Re=1$, $\sigma_r=0.1$, $\epsilon_r=2$, $\beta_i = 1/9$, $\varepsilon = 0.25$. (a) $Ca_E=0.3$ (b) $Ca_E=0.31$ (c) $Ca_E=0.32$ (d) $Ca_E=0.35$}
    \label{Fig_timeEvolution_OB_minus}
\end{figure}
\autoref{Fig_timeEvolution_OB_minus} illustrates the temporal evolution of droplet deformation and interfacial morphology for varying Deborah numbers ($De$) at four distinct electric capillary numbers, $Ca_E=0.3,0.31,0.32,0.35$ for the pair of $(\sigma_r,\epsilon_r) = (0.1, 2)$ from $OB^-$ region. At $Ca_E=0.3$, stable oblate droplet configurations are observed for $De \geq 2.5$, while instability and breakup occur for lower Deborah numbers ($De=0,0.5,1$). The characteristic breakup time increases with $De$ in the unstable regime, indicating viscoelastic retardation of the breakup. The deformation versus dimensionless time curves, truncated at the point of breakup, exhibit a sharp decrease in deformation immediately preceding the breakup, thus defining the breakup time. For $Ca_E=0.31$, drop breakup is observed across the investigated $De$ range. For $De \geq 0.5$, the breakup time increases with $De$, while the breakup time for $De=0$ is marginally greater than that for $De=0.5$. A similar trend is evident at $Ca_E=0.32$, with increasing breakup times with $De$ for $De \geq 0.5$ and an increase in breakup time as $De$ decreases from 0.5 to 0. Notably, the breakup times for $De=0$ and $De=1$ are comparable.
At the highest investigated electric Capillary number, $Ca_E=0.35$, a similar dependence of breakup time on $De$ is observed, with the breakup time generally increasing with $De$ for $De \geq 0.5$. Furthermore, the breakup time for $De=0$ is greater than that for $De=1$. We also note that the decrease in breakup time for a fixed $De$ with increasing $Ca_E$. 

\begin{figure}[ht!]
    \centering
    \includegraphics[width=1\textwidth]{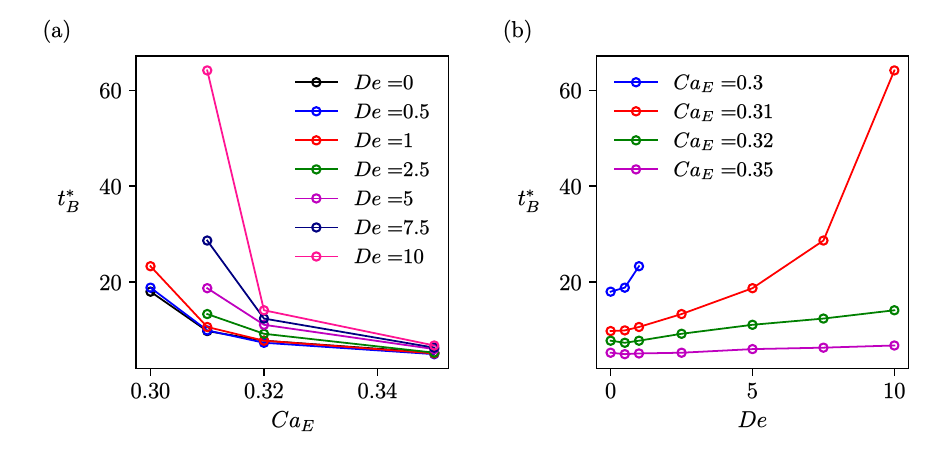}
    \caption{The variation of breakup time with (a) $Ca_E$ and (b) $De$ for $\rho_r=1$, $\mu_r=1$, $Re=1$, $\sigma_r=0.1$, $\epsilon_r=2$, $\beta_i = 1/9$, $\varepsilon = 0.25$.}
    \label{Fig_breakTimes_OB_minus}
\end{figure}
To elucidate the dependence of the dimensionless droplet breakup time on the electric Capillary number ($Ca_E$) and the Deborah number ($De$), 
\autoref{Fig_breakTimes_OB_minus}(a) and (b) illustrate the respective variations. \autoref{Fig_breakTimes_OB_minus}(a) reveals a monotonic inverse relationship between the breakup time and $Ca_E$ across all examined $De$. Conversely, \autoref{Fig_breakTimes_OB_minus}(b) demonstrates a monotonic increase in breakup time with $De$ at $Ca_E=0.3$. However, for $Ca_E \geq 0.31$, a slight decrease is observed in the breakup time as $De$ increases from 0 to 0.5, followed by a monotonic increase in breakup time with further increments in $De$. 
Notably, the maximum investigated Deborah number ($De=10$) exhibits a significantly higher rate of decrease in breakup time with increasing electric capillary number ($Ca_E$) compared to lower $De$ values. Consequently, at the maximum investigated $Ca_E=0.35$, the breakup times across the different Deborah numbers converge to comparable values, indicating a diminishing influence of viscoelasticity on the breakup dynamics at higher electric field strengths.

\subsection{Comparison with Oldroyd-B drops}
The Oldroyd-B and Linear Phan-Thien-Tanner (LPTT) models differ fundamentally in how they predict stress growth under extension. 
In the Oldroyd-B model, the polymer stress evolves linearly with the deformation rate, which causes the extensional viscosity  to diverge at a finite rate of extension, reflecting the unbounded stress growth associated with the assumption of infinitely extensible Hookean springs.
In contrast, the LPTT model modifies the constitutive equation by introducing a nonlinear damping function,
\begin{equation}
    f(\trace\,\boldsymbol{\tau}_p) = 1 + \varepsilon \frac{\lambda}{\eta_p}\,\trace\bm{\tau}_p,
\end{equation}
which effectively limits the increase of stress at large deformation rates.
Consequently, while Oldroyd-B predicts unphysical divergence in stress and extensional viscosity, the LPTT model exhibits strain-hardening at moderate $De$ and self-limiting stress at high $De$, representing the finite extensibility of polymer chains and offering a more realistic description of viscoelastic fluid behavior under strong extensional or electric stresses.

\begin{figure}[ht!]
    \centering
    \includegraphics{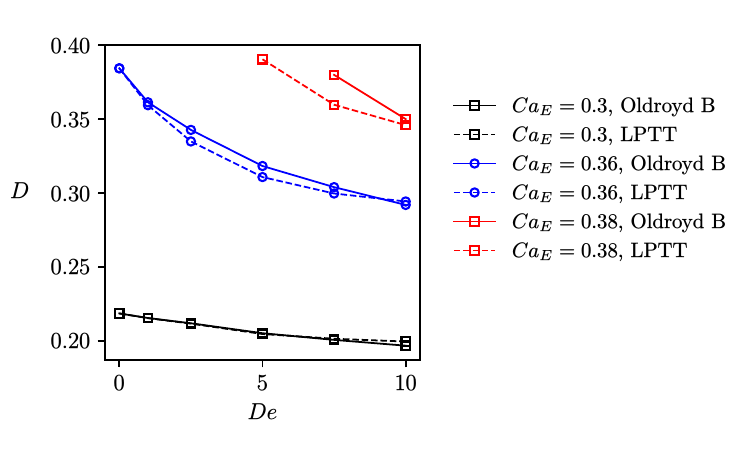}
    \caption{Deformation versus $De$ for various $Ca_E$, for $\rho_r=1$, $\mu_r=1$, $Re=1$, $\sigma_r=10$, $\epsilon_r=1.37$, $\beta_i = 1/9$, $\varepsilon=0.25$ for LPTT model}
    \label{Fig_comparison_deformation_PRA_plus}
\end{figure}
In the $PR_A^+$ region, for $(\sigma_r, \epsilon_r) = (10, 1.37)$, both Oldroyd-B (refer to \cite{bangar2026large}) and LPTT drops deform into prolate shapes, with internal flow directed from the equator to the poles.
The overall electrohydrodynamic behavior of Oldroyd-B and LPTT drops exhibits notable similarities with a few key differences when $(\sigma_r, \epsilon_r) = (10, 1.37)$. Both fluid types demonstrate a critical electric capillary number ($Ca_E$), below which the drop deforms to a stable spheroidal shape. A crucial distinction arises beyond this critical $Ca_E$: Oldroyd-B drops consistently deform into stable multi-lobed shapes across all Deborah numbers ($De$) investigated, as discussed in \cite{bangar2026large}. Conversely, LPTT drops form stable multi-lobed shapes only at low $De$ values ($De=0,1,2.5$), while they tend to break up at higher $De$ ($De=5,7.5,10$). An exception to this LPTT breakup at high $De$ is observed at $Ca_E=0.42$ and $De=5$, where the drop recovers from necking to form a stable two-lobed shape. For the both Oldroyd-B and LPTT models, drop deformation decreases monotonically with increasing $De$.
\autoref{Fig_comparison_deformation_PRA_plus} provides a detailed comparison of deformation versus $De$ for Oldroyd-B and LPTT drops at three distinct $Ca_E$ values: 0.3, 0.36, and 0.38. This analysis specifically focuses on conditions where both fluid types form stable spheroidal shapes.
At $Ca_E=0.3$, the deformation vs. $De$ curves for both Oldroyd-B and LPTT drops show nearly overlapping behavior for $De \leq 7.5$. However, at $De=10$, the LPTT drop exhibits a very slightly higher deformation compared to the Oldroyd-B drop. Moving to $Ca_E=0.36$, the deformations are almost identical for both fluids at both low and high $De$ values. For intermediate $De$ values ($De=2.5, 5, 7.5$), the deformation of LPTT drop is marginally lower.
Here, both electric and hydrodynamic stresses elongate the drop, while elastic stresses resist stretching, leading to decreasing deformation with increasing $De$. Because electric stresses are relatively weak and uniform for this set of conductivity and permittivity ratios, non-linear damping in the LPTT model is minor, resulting in similar deformation trends for both models.

\begin{figure}[ht!]
    \centering
    \includegraphics{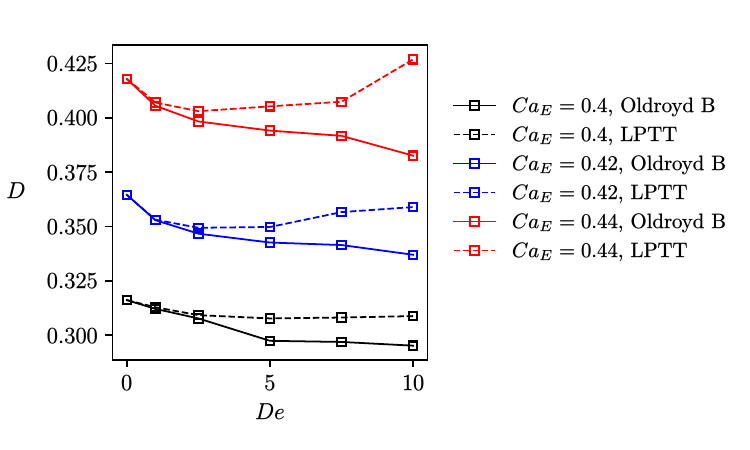}
    \caption{Variation of deformation parameter with $De$ for various $Ca_E$ for $\rho_r=1$, $\mu_r=1$, $Re=1$, $\sigma_r=25$, $\epsilon_r=50$, $\beta_i = 1/9$}
    \label{Fig_comparison_PRB_plus_1}
\end{figure}
\begin{figure}[ht!]
    \centering
    \includegraphics{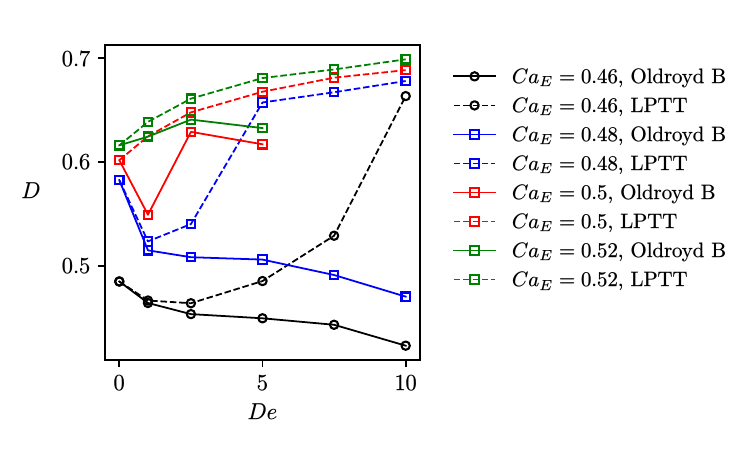}
    \caption{Variation of deformation parameter with $De$ for various $Ca_E$ for $\rho_r=1$, $\mu_r=1$, $Re=1$, $\sigma_r=25$, $\epsilon_r=50$, $\beta_i = 1/9$}
    \label{Fig_comparison_PRB_plus_2}
\end{figure}
In the $PR_B^+$ region, for $(\sigma_r, \epsilon_r) = (25, 50)$, the drop assumes a prolate shape, but the internal flow reverses direction and it is from the poles toward the equator. Here, the hydrodynamic stress opposes the electric stress, yet the latter dominates, driving prolate deformation.
When comparing the behavior of Oldroyd-B and LPTT drops under a steady electric field, distinct differences in their time evolution are evident. the deformation of an LPTT drop monotonically increases with time until it reaches a steady-state value.
In contrast, for higher Deborah numbers ($De$), the deformation of an Oldroyd-B drop (see: \cite{bangar2026large}) typically exhibits a more complex transient behavior. It often passes through a maximum deformation before settling into a steady state, sometimes even undergoing a subsequent minimum. For example, at $Ca_E=0.48$, the drop deformation shows both a maximum and a minimum before reaching its steady-state value, even though the drop shape consistently remains spheroidal.
Furthermore, in certain scenarios, the Oldroyd-B drop can transiently deform into a pointed shape during its maximum deformation phase, subsequently returning to a spheroidal steady state after passing through a minimum. This is observed for $De=10$ at $Ca_E=0.5$ and $Ca_E=0.52$. In some cases, such as for $De=7.5$ at $Ca_E=0.5$ and $0.52$, the drop may even oscillate between pointed and spheroidal shapes without achieving a stable steady state.
To compare the influence of the Deborah number ($De$) on the electrohydrodynamic deformation of Oldroyd-B and LPTT drops, \autoref{Fig_comparison_PRB_plus_1} and \autoref{Fig_comparison_PRB_plus_2} illustrate the variation of steady-state deformation with $De$ for both fluid types across various electric capillary numbers ($Ca_E$).
\autoref{Fig_comparison_PRB_plus_1} reveals that for moderate $Ca_E$ values ($0.4, 0.42, 0.44$), the steady-state deformation of an Oldroyd-B drop is consistently lower than that of an LPTT drop across all non-zero $De$ values examined. A key distinction in their deformation behavior is observed: the LPTT drop exhibits non-monotonic deformation with respect to $De$, whereas the Oldroyd-B drop's deformation monotonically decreases with increasing $De$. Furthermore, as $Ca_E$ increases, the difference in steady-state deformation between the two fluid models widens. These differences arise from the stronger and more localized electric stresses near the poles in this high-conductivity, high-permittivity region $PR_B^+$. The intense extension activates nonlinear stress responses in the LPTT model, initial strain hardening at moderate $De$, followed by stress saturation at large $De$. Consequently, LPTT drops exhibit a non-monotonic dependence of deformation on $De$, while Oldroyd-B drops display a steady decline in deformation with increasing elasticity.
\autoref{Fig_comparison_PRB_plus_2} extends this comparison to higher $Ca_E$ values ($0.46, 0.48, 0.5, 0.52$). At $Ca_E=0.46$, the gap in deformation between the two models further increases, with the LPTT drop deforming to a pointed shape at $De=10$, while the Oldroyd-B drop maintains a spheroidal steady shape. As $Ca_E$ increases to 0.48, the LPTT drop forms pointed shapes for $De \geq 5$ and also for $De=0$ (corresponding to Newtonian behavior), while the Oldroyd-B drop continues to deform into stable, spheroidal steady shapes for all non-zero $De$. For $Ca_E \geq 0.5$, the LPTT drop consistently deforms into a pointed shape across all $De$. Conversely, the Oldroyd-B drop achieves pointed steady shapes for all $De \leq 5$ (with an exception at $De=1$ for $Ca_E=0.5$), and for all $De \leq 5$ at $Ca_E=0.52$.
Examining the conditions ($Ca_E$ and $De$) under which drops deform into pointed steady shapes, we observe that for LPTT drops, deformation consistently increases with $De$. However, for Oldroyd-B drops, the deformation in pointed shapes exhibits non-monotonic dependence on $De$: it increases as $De$ rises from 0 to 2.5, then decreases as $De$ further increases to 5. Interestingly, for Oldroyd-B fluids at $De \geq 7.5$, steady pointed shapes are not achieved. Instead, the drop either oscillates between pointed and spheroidal shapes or transiently forms a pointed shape before settling into a spheroidal steady state as observed by \cite{bangar2026large}.

\begin{figure}[ht!]
    \centering
    \includegraphics{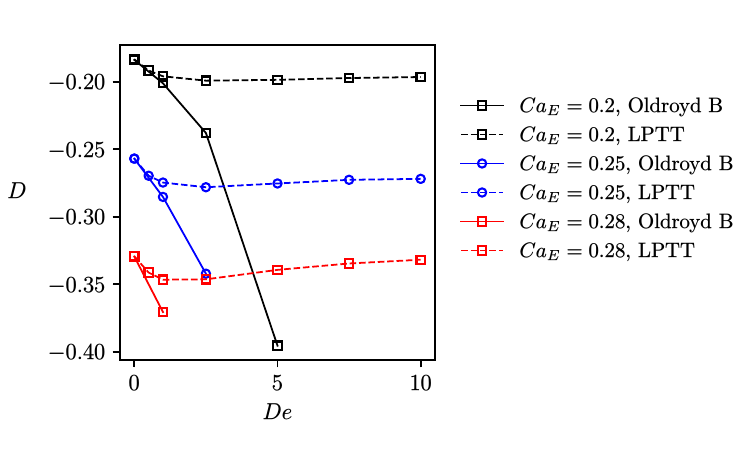}
    \caption{(a) Deformation versus $Ca_E$ for various $De$, (b) Deformation versus $De$ for various $Ca_E$, for $\rho_r=1$, $\mu_r=1$, $Re=1$, $\sigma_r=0.1$, $\epsilon_r=2$, $\beta_i = 1/9$, $\varepsilon=0.25$ for LPTT model. (from $OB^-$)}
    \label{Fig_comparison_deformation_OB_minus}
\end{figure}
In the $OB^-$ region, the drop attains an oblate shape due to the electric field (for the selected conductivity ratio and permittivity ratio electric field tends to deform drop to oblate shape as described by \cite{esmaeeli2020transient}), with the flow directed from the poles toward the equator. 
Comparing the behavior of Oldroyd-B drops (discussed in \cite{bangar2026large}) with LPTT drops in a steady electric field reveals significant differences, even with some shared characteristics. Consistent with Newtonian fluid behavior, both Oldroyd-B and LPTT drops exhibit a critical electric capillary number ($Ca_E$). Below this value, the drop can deform into a stable shape; exceeding it leads to breakup.
However, their stable deformation modes differ. LPTT drops deform to a stable spheroidal shape when $Ca_E$ is below its critical value. In contrast, Oldroyd-B drops, while mainly forming stable spheroidal shapes below critical $Ca_E$, can also form stable shapes with dimples ($Ca_E=0.2, De=5$). In some instances (specifically at low $Ca_E$ and high $De$, for example, $De \geq 7.5$ for $Ca_E=0.1$ and 0.2), Oldroyd-B drops can even form dimples and attempt to break up, only to recover into a spheroidal shape and undergo positional oscillations.
Furthermore, LPTT drops generally demonstrate greater stability, showing a higher critical $Ca_E$ at elevated $De$ values. Oldroyd-B drops, on the contrary, show a lower critical $Ca_E$ for higher $De$. It is important to note that the LPTT drops do not breakup for any $De$ up to $Ca_E=0.28$. However, Oldroyd-B drops experience breakup for $De \geq 2.5$ at $Ca_E=0.28$, and for $Ca_E=0.25$, breakup occurs at $De \geq 5$. Although Oldroyd-B drops do not break up at $Ca_E=0.2$ for any $De$, for $De \geq 7.5$, they attempt to break and then recover, leading to positional oscillations. Regarding breakup kinetics, LPTT drops generally take longer to break up at higher $De$, whereas Oldroyd-B drops break up more quickly under similar conditions.
\autoref{Fig_comparison_deformation_OB_minus} illustrates the variation of the deformation parameter with $De$ for Oldroyd-B and LPTT drops at three values of $Ca_E$: 0.2, 0.25, and 0.28. The deformation parameter (magnitude of deformation) for LPTT drops exhibits non-monotonic behavior, initially decreasing (increasing) and then increasing (decreasing) with $De$. Conversely, for Oldroyd-B drops, it monotonically decreases (increases) with $De$. The magnitude of deformation is significantly higher for Oldroyd-B drops compared to LPTT drops. Thus, for Oldroyd-B drops, increasing $De$ generally enhances drop deformation and leads to earlier breakup. In contrast, LPTT drops display a non-monotonic response where the magnitude of deformation initially increases due to strain hardening and then decreases at larger $De$ owing to stress saturation, indicating greater resistance to deformation and breakup at high $De$ values.

\section{Conclusion}\label{Sec_Conclusion}
The deformation and breakup dynamics of a linear PTT drop under an electric field are numerically investigated and compared with those of Oldroyd-B drops. Representative pairs of conductivity ($\sigma_r$) and permittivity ($\epsilon_r$) ratios are selected from six regions of the ($\sigma_r$, $\epsilon_r$) phase plot, $PR_A^+$, $PR_B^+$, $PR_A^-$, $PR_B^-$, $OB^-$, and $OB^+$.
For $(\sigma_r, \epsilon_r)$ pairs within $PR_A^-$, $PR_B^-$, and $OB^+$ regions, drop deformation increases gradually with the electric capillary number ($Ca_E$), leading to smooth shape evolution and stable steady states even at high $Ca_E$. In these regions, viscoelastic effects are minimal.

For $(\sigma_r, \epsilon_r)$ pair within the $PR_A^+$ region, an LPTT drop deforms into a stable spheroidal shape when the electric capillary number ($Ca_E$) is below a critical value. Beyond this critical $Ca_E$, the drop forms stable multi-lobed shapes or breaks up. Drop deformation decreases with increasing Deborah number ($De$), indicating an enhanced resistance to deformation with increasing viscoelasticity. The overall behavior closely resembles that of an Oldroyd-B drop, with some notable distinctions. At both low and high $De$, the deformation of Oldroyd-B and LPTT drops is nearly identical. However, for intermediate $De$ values, Oldroyd-B drops exhibit slightly higher deformation than LPTT drops. Furthermore, the critical $Ca_E$ for Oldroyd-B drops is observed to be slightly lower than or similar to that of LPTT drops, depending on $De$.

In the $PR_B^+$ region, LPTT drops deform into a steady spheroidal shape up to a critical $Ca_E$ and forms shapes with pointed ends when critical $Ca_E$ is exceeded. We found that, when $Ca_E$ is below critical threshold, contrary to Oldroyd-B drops, the deformation of an LPTT drop shows non-monotonic behavior with $De$, it increases with $De$ initially followed by a decrease with increasing $De$.
Specifically, both $De=0$ (Newtonian behavior) and very high $De$ values result in pointed shapes at lower $Ca_E$. In contrast, intermediate $De$ values require a comparatively higher $Ca_E$ to induce pointed shapes, implying that the critical $Ca_E$ first increases with $De$ before subsequently decreasing. This indicates that in the $PR_B^+$ region, for intermediate $De$, increased viscoelasticity provides greater resistance to deformation, while for sufficiently large $De$, this resistance diminishes, and deformation is enhanced compared to a Newtonian drop.
Overall, LPTT drops exhibit larger deformation compared to Oldroyd-B drops for all non-zero $De$ values, indicating a lesser resistance to deformation as compared to Oldroyd-B drop.

For ($\sigma_r, \epsilon_r$) from the $OB^-$ region, an LPTT drop deforms into a stable oblate spheroidal shape below the critical $Ca_E$ and drop breaks up above it. 
In stark contrast to Oldroyd-B drops, LPTT drops exhibit a non-monotonic relationship with $De$, where magnitude of deformation initially increases before subsequently decreasing, ultimately leading to the lower deformation at higher $De$. This fundamental difference suggests that at elevated $De$ values, Oldroyd-B drops experience enhanced deformation and are more prone to breakup, while LPTT drops demonstrate greater resistance to deformation and tend to be more stable.




\bibliographystyle{elsarticle-harv} 
\bibliography{references.bib}



\end{document}